\documentclass{jpp}
\usepackage{graphicx}
\usepackage{epstopdf, epsfig}
\usepackage{hyperref}
\usepackage{amsmath}
\usepackage[normalem]{ulem}
\usepackage{stackengine}
\usepackage{color}

\newcommand{\tderiv}[1]{\frac{\partial #1}{\partial t}}
\newcommand{\pgrad}{\nabla^\perp}
\newcommand{\psderiv}{\partial_{\psi_v}}
\newcommand{\llderiv}{\partial^\parallel}
\newcommand{\pLap}{\Delta^\perp}
\newcommand{\vpar}{v_\parallel}
\newcommand{\st}[1]{\hbox{\sout{$#1$}}}
\newcommand{\const}{\mathrm{const}}

\shorttitle{Testing of the stellarator-capable model}
\shortauthor{N. Nikulsin, M. Hoelzl, A. Zocco, K. Lackner, S. G\"unter and the JOREK Team}

\title{Testing of the new JOREK stellarator-capable model in the tokamak limit}

\author{Nikita Nikulsin\aff{1}
  \corresp{\email{nikita.nikulsin@ipp.mpg.de}},
  Matthias Hoelzl\aff{1},
  Alessandro Zocco\aff{2},
  Karl Lackner\aff{1},
  Sibylle G\"unter\aff{1}
  \and the JOREK Team}

\affiliation{\aff{1}Max Planck Institute for Plasma Physics, Boltzmannstr. 2, 85748 Garching, Germany
\aff{2}Max Planck Institute for Plasma Physics, Wendelsteinstr. 1, 17491 Greifswald, Germany}

\begin{document}

\maketitle

\begin{abstract}
In preparation for extending the JOREK nonlinear MHD code to stellarators, a hierarchy of stellarator-capable reduced and full MHD models has been derived and tested. The derivation was presented at the EFTC 2019 conference. Continuing this line of work, we have implemented the reduced MHD model \citep{nikulsin2019a} as well as an alternative model which was newly derived using a different set of projection operators for obtaining the scalar momentum equations from the full MHD vector momentum equation. With the new operators, the reduced model matches the standard JOREK reduced models for tokamaks in the tokamak limit and conserves energy exactly, while momentum conservation is less accurate than in the original model whenever field-aligned flow is present.
\end{abstract}

\section{Introduction}

Reduced MHD, a system of approximations introduced in its original form in the 1960s \citep{greene1961determination}, continues to be used by modern MHD codes, such as JOREK \citep{franck2015energy} and M3D-$C^1$ \citep{breslau2009some}. While many versions of reduced MHD consisting of different equations and using different methods to derive them have been published, the main idea is the same: the removal of fast magnetosonic waves while retaining relevant physics \citep{strauss1997reduced,jardin2012multiple}. The removal of these waves eliminates the shortest time scale and allows one to use larger time steps due to the Courant condition. Even when implicit time integration methods are used, and the Courant condition is no longer a hard limit, using time steps that are large compared to the shortest time scale can lead to poor accuracy \citep{jardin2012multiple,kruger1998generalized}. An increased time step is especially advantageous in nonlinear simulations, which otherwise tend to be computationally demanding. In particular, stellarator simulations, which are even more computationally expensive than tokamak simulations due to the complex geometry, can benefit from an increased step size.

Originally, reduced MHD was derived via an ordering in a small parameter, often taken to be the inverse aspect ratio. The ordering itself is a system of several approximations and assumptions involving the ordering parameter that allows one to determine the relative order (in terms of the ordering parameter) of any quantity with respect to any other quantity of the same dimension. In this context, terms corresponding to fast magnetosonic waves have a higher order than the terms that one wants to keep, allowing them to be dropped. Different versions of reduced MHD can be derived by using different orderings that keep different physical effects (e.g. low-$\beta$ vs high-$\beta$ orderings) \citep{kruger1998generalized,strauss1976nonlinear,strauss1977dynamics,strauss1980stellarator,strauss1997reduced}. An alternative ansatz-based approach was introduced by \citet{park1980non}, where an ansatz form that eliminates fast magnetosonic waves is used for the velocity and terms of all orders are kept. Later papers also adopt an ansatz for the magnetic field that eliminates field compression \citep{huysmans2007mhd,breslau2009some,franck2015energy}. The ansatz-based approach makes less assumptions and keeps more physical effects, while generally resulting in more complicated equations.

An ordering-based reduced MHD model for stellarators was recently derived in \citet{zocco2020magnetic} with $\beta\sim\epsilon\ll 1$, where $\epsilon$ is the inverse aspect ratio. Here, as in our previous paper \citep{nikulsin2019a}, we adopt an ansatz-based approach. Previously, we derived a hierarchy of models, which we intended to use for stellarator simulations. We started by introducing ansatzes for the magnetic field and velocity and proving that any arbitrary magnetic field and velocity can be represented by those ansatzes. We also showed that if the background vacuum field is strong enough, then the three terms of velocity ansatz approximately separate the three MHD waves. Dropping the fast magnetosonic wave term from the velocity ansatz and the field compression term from the magnetic field ansatz, we obtain reduced MHD \citep{nikulsin2019a}. However, we later found that the projection operators used had good momentum conservation properties, but not as good energy conservation properties. We thus adopt a different set of projection operators, which makes our new stellarator models a direct generalization of the currently used JOREK tokamak models to three dimensional geometries; the stellarator models reduce back to the tokamak models in the tokamak limit.

The rest of this paper is structured as follows. In section \ref{sec:changes}, we briefly review the original set of models, introduce the changes that we made and derive the new set of models. In section \ref{sec:encon}, we show how the deficiencies of our original model are fixed in the new model. In section \ref{sec:examples}, we provide some test runs, which show how the deficiencies of the original model manifest themselves in practice by comparing it to the new stellarator model/standard tokamak model (both are the same in this situation) in the case of a tearing mode in a tokamak. In section \ref{sec:momentum}, we consider analytically the local momentum conservation properties of both the original and new models, and then compare the global momentum conservation of the new stellarator model with and without field-aligned flow in the case of a ballooning mode. Finally, in Appendix \ref{sec:appendix}, a technicality in the implementation of the original model is discussed.

\section{Changes made to the model}\label{sec:changes}

The original hierarchy of models, as presented in \citet{nikulsin2019a}, was derived from the viscoresistive MHD equations:
\begin{equation}
	\label{eq:vrmhd}
	\begin{gathered}
		\tderiv{\rho} + \nabla\cdot (\rho\vec v) = P, \\
		\begin{aligned}
		    \rho\tderiv{\vec v} + \rho(\vec v\cdot\nabla)\vec v + \vec v P &= \vec j\times\vec B - \nabla p + \mu\Delta\vec v, & \qquad \tderiv{\vec B} &= -\nabla\times\vec E,
		\end{aligned} \\
		\tderiv{\mathcal{E}} + \nabla\cdot\left[\left(\frac{\rho v^2}{2} + \frac{\gamma p}{\gamma-1}\right)\vec v + \frac{p}{\gamma-1}\frac{D_\perp}{\rho}\nabla_\perp\rho + \frac{\vec E\times\vec B}{\mu_0} - \kappa_\perp \nabla_\perp T - \kappa_\parallel \nabla_\parallel T\right] = S_e - \frac{v^2 P}{2}, \\
		\mathcal{E} = \frac{\rho v^2}{2} + \frac{p}{\gamma-1} + \frac{B^2}{2\mu_0}, \\
		\begin{aligned}
			\nabla\times\vec B &= \mu_0 \vec j, & \nabla\cdot\vec B &= 0, & \vec E &= -\vec v\times\vec B + \eta\vec j, & P = \nabla\cdot (D_\perp \nabla_\perp \rho) + S_\rho.
		\end{aligned}
	\end{gathered}
\end{equation}
The gradient operators parallel and perpendicular to the total magnetic field \(\vec B\) are defined as \(\nabla_\parallel = \frac{\vec B}{B^2}\vec B\cdot\nabla\) and \(\nabla_\perp = \nabla - \nabla_\parallel\). The usual MHD notation is followed with \(\rho\), \(p\), \(\vec v\) and \(\vec B\) being density, pressure, velocity and magnetic field, respectively. In addition to that, \(\eta\) is the resistivity, \(\mu\) is the dynamic viscosity, \(D_\perp\) is the mass diffusion coefficient perpendicular to field lines, \(\kappa_\perp\) and \(\kappa_\parallel\) are the thermal conductivity across and along field lines, and \(S_\rho\) and \(S_e\) are source terms in the continuity and energy equations, respectively. The ideal gas law \(p = \rho RT\) is assumed to hold.

We also introduced the following ansatzes for the magnetic field and velocity:
\begin{equation}
	\label{eq:anz}
	\begin{gathered}
		\vec B = \nabla\chi + \nabla\Psi\times\nabla\chi + \nabla\Omega\times\nabla\psi_v, \\
		\vec v = \frac{\nabla\Phi\times\nabla\chi}{B_v^2} + \vpar\vec B + \pgrad\zeta,
	\end{gathered}
\end{equation}
and proved that any magnetic field and any velocity can be represented in such a way. Here, $\nabla\chi$ is the curl-free vacuum component of the magnetic field, which is generated by the coils, $B_v = |\nabla\chi|$ and $\pgrad = \nabla - B_v^{-2}\nabla\chi(\nabla\chi\cdot\nabla)$. Note that, since the ansatzes do not restrict the magnetic and velocity fields, they can still be used within the context of full MHD. The transition to reduced MHD is done by setting $\Omega = \zeta = 0$ and dropping the evolution equations for $\Omega$ and $\zeta$ (derived below). Further, in the tokamak limit, i.e. when $\chi = F_0\phi$, where $\phi$ is the toroidal angle \footnote{In many tokamaks, the vacuum field also includes a contribution from the poloidal field coils, however the toroidal field is by far the dominant component of the vacuum field, and it makes sense to include only the toroidal field in $\nabla\chi$. Throughout this paper, we will refer to $\chi = F_0\phi$ as the tokamak limit.} , the ansatzes reduce to those of the standard JOREK reduced MHD tokamak models, $\vec B = F_0\nabla\phi + \nabla\psi\times\nabla\phi$ and $\vec v = R^2\nabla u\times\nabla\phi + \vpar\vec B$, where $\psi = F_0\Psi$ and $u = \Phi/F_0$ \citep{hoelzl2020the}.

The scalar $\psi_v$ is one of the Clebsch potentials for the vacuum field, which can locally be written as
\begin{equation}
    \label{eq:clebsch}
    \nabla\chi = \nabla\psi_v\times\nabla\beta_v.
\end{equation}
As we discussed in \citet{nikulsin2019a}, one can construct a Clebsch-type coordinate system with the coordinates $(\psi_v,\beta_v,\chi)$. Note that the variables $\psi_v$ and $\beta_v$ are not unique, as can be seen by replacing $\psi_v$ with $\psi_v' = \psi_v + f(\beta_v)$, where $f$ is an arbitrary function, which leaves equation \eqref{eq:clebsch} unchanged. In general, given some vacuum magnetic field $\nabla\chi$, the scalars $\psi_v$ and $\beta_v$ are completely determined by fixing some parameterization $(\psi_v,\widetilde{\beta}_v)$ of a surface intersected by the vacuum field lines, with the values of $\psi_v$ and $\widetilde{\beta}_v$ in the rest of space being determined by following the field lines off of the surface while keeping $\psi_v$ and $\widetilde{\beta}_v$ constant. In a toroidal system, the surface is usually chosen to be a poloidal plane, and, in many cases, a cut needs to be introduced at that poloidal plane to prevent multivaluedness of $\psi_v$ and $\widetilde{\beta}_v$ when the field lines travel once around the torus and encounter previously labelled points. Finally, $\beta_v$ is determined by integrating $\partial\beta_v/\partial\widetilde{\beta}_v = B_v/|\nabla\psi_v\times\nabla\widetilde{\beta}_v|$ \citep{d2012flux,stern1970euler}. Usually, it is advantageous to find a $\psi_v$ such that the component of $\nabla\Omega\times\nabla\psi_v$ perpendicular to $\nabla\chi$, $F_v^2|\llderiv\Omega|^2$, is minimized at $t = 0$, so that most of field line bending is captured by the $\nabla\Psi\times\nabla\chi$ term.

The variables $\Psi$, $\Omega$, $\Phi$, $\vpar$ and $\zeta$ are the unknown magnetic field and velocity variables, which are solved for. In order to obtain evolution equations for these variables, we projected Faraday's law onto $\nabla\psi_v$ and $\nabla\chi$, and applied the following projection operators to the vector momentum equation, which was first divided by $\rho$
\begin{equation}
	\label{eq:oldprojop}
	\begin{gathered}
		\nabla\chi\cdot\nabla\times[\nabla\chi\times(\vec e_\chi\times \\
		\nabla\chi\cdot \\
		\nabla\cdot [B_v^2 \nabla\chi\times(\vec e_\chi\times
	\end{gathered}
\end{equation}
Here, $\vec e_\chi = \vec B/B^\chi$ is the covariant basis vector in the Clebsch-type coordinate system $(\alpha,\beta,\chi)$ associated with the \emph{full} magnetic field, $\vec B = \nabla\alpha\times\nabla\beta$. These projection operators have the property that each one of them is orthogonal to all but one term in the velocity ansatz \eqref{eq:anz}, i.e.
\begin{equation*}
	\begin{gathered}
		\nabla\chi\cdot\nabla\times[\nabla\chi\times(\vec e_\chi\times\vec v)] = \pLap\Phi, \\
		\nabla\chi\cdot\vec v = B^\chi \vpar, \\
		\nabla\cdot [B_v^2\nabla\chi\times(\vec e_\chi\times\vec v)] = -\nabla\cdot(B_v^2 \pgrad \zeta),
	\end{gathered}
\end{equation*}
where $\pLap = \nabla\cdot\pgrad$. However, for the sake of energy conservation, a better set of projection operators would be
\begin{equation}
	\label{eq:newprojop}
	\begin{gathered}
		\nabla\chi\cdot\nabla\times(B_v^{-2} \\
		\vec B\cdot \\
		\nabla\chi\cdot\nabla\times(B_v^{-2}\nabla\chi\times
	\end{gathered}
\end{equation}
as we will show in the next section.

To obtain the new evolution equations for the velocity variables, we insert the ansatzes \eqref{eq:anz} into the momentum equation and apply the projection operators \eqref{eq:newprojop}, this time without dividing first by $\rho$. The momentum equation can be written as
\begin{equation}
\label{eq:nsins}
\rho\nabla\tderiv{\Phi}\times\frac{\nabla\chi}{B_v^2} + \rho\tderiv{}(\vpar\vec B) + \rho\pgrad\tderiv{\zeta} + \frac{\rho}{2}\nabla v^2 + \rho\vec\omega\times\vec v + \vec v P = \vec j\times\vec B - \nabla p,
\end{equation}
where we used the identity \((\vec v\cdot\nabla)\vec v = \frac{1}{2}\nabla v^2 + \vec\omega\times\vec v\) and \(\vec\omega\) is the vorticity:
\begin{equation}
\vec\omega = \nabla\times\vec v = -\nabla\chi\nabla\cdot\left(\frac{\nabla\Phi}{B_v^2}\right) + B_v\llderiv\frac{\nabla\Phi}{B_v^2} - \frac{\nabla\Phi\cdot\nabla}{B_v^2}\nabla\chi + \nabla \vpar\times\vec B + \mu_0 \vpar\vec j + \nabla\chi\times\nabla\frac{\llderiv\zeta}{B_v}.
\end{equation}
Just as in \citet{nikulsin2019a}, we do not carry the viscosity term through this derivation, instead adding a generic viscosity term afterwards. To obtain the evolution equation for $\Phi$, we apply to equation \eqref{eq:nsins} the first projection operator in \eqref{eq:newprojop}, or its equivalent, $-\nabla\cdot(B_v^{-2}\nabla\chi\times$, when appropriate:
\begin{equation}
	\label{eq:phieq}
	\begin{aligned}
		&\nabla\cdot\left[\frac{\rho}{B_v^2}\pgrad\tderiv{\Phi} + \rho\tderiv{}(\vpar\pgrad\Psi) - \frac{\rho}{B_v}\tderiv{}(\vpar\llderiv\Omega)\nabla\psi_v\right] - B_v\left[\frac{\rho}{B_v^2},\tderiv{\zeta}\right] = \frac{B_v}{2}\left[\frac{\rho}{B_v^2},v^2\right] \\
		&+ B_v\left[\frac{\rho\omega^\chi}{B_v^4},\Phi\right] - \nabla\cdot(\rho \vpar\vec\omega^\perp) + B_v\left[\frac{\rho \vpar\omega^\chi}{B_v^2},\Psi\right] + B_v\left[\frac{\rho \vpar\omega^{\psi_v}}{B_v^2},\Omega\right] \\
		&- B_v\left[\frac{\rho \vpar\vec\omega\cdot\nabla\Omega}{B_v^2},\psi_v\right] + \nabla\cdot\left(\frac{\rho\omega^\chi}{B_v^2}\pgrad\zeta - \frac{P}{B_v^2}\pgrad\Phi - P\vpar\pgrad\Psi + \frac{P\vpar\llderiv\Omega}{B_v}\nabla\psi_v\right) \\
		&+ B_v\left[\frac{P}{B_v^2},\zeta\right] + \nabla\cdot\left(\frac{B^\chi}{B_v^2}\vec j - \frac{j^\chi}{B_v^2}\vec B\right) + B_v\left[\frac{1}{B_v^2},p\right] + \nabla\cdot(\mu_\perp\pgrad\pLap\Phi),
	\end{aligned}
\end{equation}
where $[f,g] = B_v^{-1}\nabla\chi\cdot(\nabla f\times\nabla g)$ is a Poisson bracket, $\llderiv = B_v^{-1}\nabla\chi\cdot\nabla$, superscripts indicate contravariant vector components, and, for any vector $\vec U$, $\vec U^\perp = \vec U - U^\chi\nabla\chi/B_v^2$. We have added the generic viscosity term ${\nabla\cdot(\mu_\perp\pgrad\pLap\Phi)}$, which we have chosen to match the viscosity term in \citet{franck2015energy} in the tokamak limit. As discussed in previous papers \citep{franck2015energy,nikulsin2019a}, the viscosity term $\mu\Delta\vec v$ in the momentum equation \eqref{eq:vrmhd} is only a rough approximation for the divergence of the viscous stress tensor in a plasma, and one does not loose much by using a generic viscosity term in the final equations. Note, however, that the $\mu_\perp$ in the above equation is not the physical viscosity $\mu$ in equation \eqref{eq:vrmhd}. Indeed, from dimensional analysis it follows that $\mu_\perp$ has units of $\mathrm{T^2\cdot Pa\cdot s}$. Applying scaling analysis to the term $\mu\Delta\vec v$ after inserting just the first term of the ansatz \eqref{eq:anz} for $\vec v$ and applying the first projection operator in \eqref{eq:newprojop}, we see that $\nabla\chi\cdot\nabla\times[B_v^{-2}\mu\Delta(\nabla\Phi\times\nabla\chi/B_v^2)] \sim \mu B_v^{-2}\Phi/L_\perp^4$, where $L_\perp$ is the length scale perpendicular to the magnetic field, and $L_\perp \ll L_\parallel$ for most magnetic fusion configurations. Applying the same analysis to the generic viscosity term, we get $\nabla\cdot(\mu_\perp\pgrad\pLap\Phi) \sim \mu_\perp\Phi/L_\perp^4$. Comparing the two scalings, we see that $\mu_\perp$ scales as $\mu B_v^{-2}$; for the purposes of the scaling, one can take the value of $B_v$ at the axis as a typical value and write $\mu_\perp \sim \mu B_{v,axis}^{-2}$.

To get the evolution equation for $\vpar$, we apply the second operator in \eqref{eq:newprojop}, projecting equation \eqref{eq:nsins} on $\vec B$:
\begin{equation}
	\label{eq:vpeq}
	\begin{aligned}
		&\rho\left(\tderiv{\Phi},\Psi\right) - \rho\frac{F_v}{B_v}\llderiv\Omega\psderiv\tderiv{\Phi} + \rho B^2\tderiv{\vpar} + \frac{\rho \vpar}{2}\tderiv{B^2} + \rho B_v\left[\tderiv{\zeta},\Psi\right] + \rho F_v\left[\tderiv{\zeta},\Omega\right]_{\psi_v} \\
		&= -\frac{\rho B_v}{2}\llderiv v^2 - \frac{\rho B_v}{2}\left[v^2,\Psi\right] - \frac{\rho F_v}{2}\left[v^2,\Omega\right]_{\psi_v} - \frac{\rho\omega^\chi}{B_v}\llderiv\Phi - \frac{\rho\omega^\chi}{B_v}\left[\Phi,\Psi\right] - \frac{\rho\omega^\chi F_v}{B_v^2}\left[\Phi,\Omega\right]_{\psi_v} \\
		&+ \frac{\rho B^\chi}{B_v^2}\vec\omega\cdot\nabla\Phi - \rho\vec\omega\cdot(\nabla\zeta\times\nabla\chi) + \rho\omega^\chi(\Psi,\zeta) - \rho\vec\omega\cdot\nabla\Omega F_v\psderiv\zeta + \rho\omega^{\psi_v}(\Omega,\zeta) - \vec v\cdot\vec B~P \\
		&- B_v\llderiv p - B_v\left[p,\Psi\right] - F_v\left[p,\Omega\right]_{\psi_v} + \nabla\cdot(\mu_\parallel\pgrad \vpar),
	\end{aligned}
\end{equation}
where we have again added a generic viscosity term with a form analogous to the viscosity term in the previous equation. Here, $(f,g) = \pgrad f\cdot\pgrad g$ is an inner product, $F_v = |\nabla\psi_v|$, $\psderiv = F_v^{-1}\nabla\psi_v\cdot\nabla$ and $[f,g]_{\psi_v} = F_v^{-1}(\nabla f\times\nabla g)$ is a Poisson bracket with respect to $\nabla\psi_v$. Finally, to obtain the evolution equation for $\zeta$, we apply the last projection operator in \eqref{eq:newprojop} (or its equivalent, ${-\nabla\cdot[B_v^{-2}\nabla\chi\times(\nabla\chi\times}$, to some terms) to equation \eqref{eq:nsins}:
\begin{equation}
	\label{eq:zetaeq}
	\begin{aligned}
		&B_v\left[\frac{\rho}{B_v^2},\tderiv{\Phi}\right] + B_v\left[\rho\tderiv{\vpar},\Psi\right] + B_v\left[\rho \vpar,\tderiv{\Psi}\right] - B_v\left[\frac{\rho}{B_v}\tderiv{}(\vpar\llderiv\Omega),\psi_v\right] \\
		&+ \nabla\cdot\left(\rho\pgrad\tderiv{\zeta}\right) = -\nabla\cdot\left(\frac{\rho}{2}\pgrad v^2 + \frac{\rho\omega^\chi}{B_v^2}\pgrad\Phi - \frac{\rho \vpar B^\chi}{B_v^2}\nabla\chi\times\vec\omega + \frac{\rho \vpar\omega^\chi}{B_v^2}\nabla\chi\times\vec B\right) \\
		&+ B_v\left[\frac{\rho\omega^\chi}{B_v^2},\zeta\right] - B_v\left[\frac{P}{B_v^2},\Phi\right] - B_v\left[P\vpar,\Psi\right] + B_v\left[\frac{P\vpar\llderiv\Omega}{B_v},\psi_v\right] \\
		&- \nabla\cdot\left(P\pgrad\zeta + \frac{B^\chi}{B_v^2}\nabla\chi\times\vec j - \frac{j^\chi}{B_v^2}\nabla\chi\times\vec B\right) - \pLap p + \nabla\cdot(\mu_\zeta\pgrad\pLap\zeta).
	\end{aligned}
\end{equation}
This concludes the derivation of the scalar momentum equations.

The evolution equations for the magnetic field variables of \citet{nikulsin2019a} are valid, therefore we keep them as is:
\begin{equation}
	\label{eq:mfeqns}
	\begin{aligned}
		\left[\psi_v,\tderiv{\Psi}\right] &= \left[\frac{[\Psi,\Phi]-\llderiv\Phi}{B_v},\psi_v\right] - \frac{F_v}{B_v}\left[\Omega,\frac{[\psi_v,\Phi]}{B_v}\right]_{\psi_v} + \llderiv(F_v\psderiv\zeta) + \left[(\zeta,\Psi),\psi_v\right] \\
		&- \frac{F_v}{B_v}\left[\Omega,(\zeta,\psi_v)\right]_{\psi_v} + \frac{1}{B_v}\nabla\cdot(\eta\nabla\psi_v\times\vec j), \\
		\left[\tderiv{\Omega},\psi_v\right] &= -\left[\Omega,\frac{[\psi_v,\Phi]}{B_v}\right] + \left[\psi_v,\frac{[\Omega,\Phi]}{B_v}\right] - 2(B_v,\zeta) - B_v\pLap\zeta - \left[\Omega,(\zeta,\psi_v)\right] \\
		&+ \left[\psi_v,(\zeta,\Omega)\right] + \frac{1}{B_v}\nabla\cdot(\eta\nabla\chi\times\vec j).
	\end{aligned}
\end{equation}
The continuity equation remains unchanged \citep{nikulsin2019a}:
\begin{equation}
	\label{eq:rhoeq}
	\tderiv{\rho} = - B_v\left[\frac{\rho}{B_v^2},\Phi\right] - B_v\llderiv(\rho \vpar) - B_v[\rho \vpar,\Psi] - F_v[\rho \vpar,\Omega]_{\psi_v} - \nabla\cdot(\rho\pgrad\zeta) + P,
\end{equation}
while for the pressure, we use the standard internal energy evolution equation, instead of the full energy conservation equation from the system \eqref{eq:vrmhd}:
\begin{equation}
	\label{eq:peq0}
	\tderiv{p} + \vec v\cdot\nabla p + \gamma p\nabla\cdot\vec v = (\gamma-1)\left[\nabla\cdot\left(\kappa_\perp\nabla_\perp T + \kappa_\parallel\nabla_\parallel T + \frac{p}{\gamma - 1}\frac{D_\perp}{\rho}\nabla_\perp\rho\right) + S_e + \eta j^2\right],
\end{equation}
which now, after using \eqref{eq:anz}, reads
\begin{equation}
	\label{eq:peq}
	\begin{aligned}
		\tderiv{p} &= - \frac{1}{B_v}\left[p,\Phi\right] - \vpar B_v\llderiv p - \vpar B_v\left[p,\Psi\right] - \vpar F_v \left[p,\Omega\right]_{\psi_v} - (\zeta,p) - \gamma p B_v\left[\frac{1}{B_v^2},\Phi\right] \\
		&- \gamma p B_v\llderiv \vpar - \gamma p B_v\left[\vpar,\Psi\right] - \gamma p F_v\left[\vpar,\Omega\right]_{\psi_v} - \gamma p\pLap\zeta + \nabla\cdot\Bigg[(\gamma-1)\frac{\kappa_\perp}{R}\nabla\left(\frac{p}{\rho}\right) \\
		&+ (\gamma-1)\frac{\kappa_\parallel-\kappa_\perp}{RB^2}\vec B\left(B_v\llderiv\left(\frac{p}{\rho}\right) + B_v\left[\frac{p}{\rho},\Psi\right] + F_v\left[\frac{p}{\rho},\Omega\right]_{\psi_v}\right) \\
		&+ \frac{p D_\perp}{\rho}\left(\nabla\rho-\frac{\vec B}{B^2}(B_v\llderiv\rho + B_v[\rho,\Psi] + F_v[\rho,\Omega]_{\psi_v})\right)\Bigg] + (\gamma-1)(S_e + \eta j^2).
	\end{aligned}
\end{equation}

We note that since any arbitrary magnetic field and velocity can be represented in the ansatz form \eqref{eq:anz}, the equations derived above still correspond to full MHD, albeit in a unconventional form. We also showed that if the background vacuum field is the dominant part of the magnetic field, then the MHD waves are approximately separated by the terms of the velocity ansatz, with the first term containing Alfv\'en waves, the second containing slow magnetosonic waves, and the last one containing fast magnetosonic waves \citep{nikulsin2019a}. Thus, to get the reduced MHD system, we set $\zeta = \Omega = 0$, eliminating fast magnetosonic waves and field compression, and drop the corresponding evolution equations, namely equation \eqref{eq:zetaeq} and the second equation in \eqref{eq:mfeqns}, to avoid having an overconstrained system. We also approximate the electric field in Ohm's Law as $\vec E = -\vec v\times\vec B + \eta\vec j^\parallel$, which corresponds to neglecting the components of $\vec j$ perpendicular to $\nabla\chi$ in the last term of the first equation in \eqref{eq:mfeqns}. Alternatively, if we introduce a tensor resistivity, $\overleftrightarrow{\eta}$, this approximation corresponds to neglecting the perpendicular resistivity, $\eta^\perp \approx 0$, although parallel and perpendicular resistivities are usually of the same order, so it makes more sense to speak about neglecting the perpendicular current (only in the resistive term). This seems to be a fairly good approximation, at least for tokamaks \citep{pfirsch1978equilibrium}, and likely also for any configuration with significant bootstrap current. Indeed, we can write $\overleftrightarrow{\eta}\cdot\vec j = \eta_\perp\vec j_\perp + \eta_\parallel\vec j_\parallel$, where subscripts refer to parallel and perpendicular components relative to the total field $\vec B$, not just $\nabla\chi$. When the bootstrap current is high enough, the $\eta_\parallel\vec j_\parallel$ term is dominant, and, since $\nabla\chi$ is the dominant component of $\vec B$, we have $\vec j_\parallel \approx \vec j^\parallel$ in the lowest order. As we will show in the next section, the approximation is necessary in order to have energy conservation in reduced MHD. Accordingly, we also have to replace the last term of equation \eqref{eq:peq} with $\eta (j^\parallel)^2$ to avoid creating extra internal energy, as now only the parallel component of current contributes to the dissipation of magnetic energy. A further reduction would be to also eliminate slow magnetosonic waves by setting $\vpar = 0$ and dropping equation \eqref{eq:vpeq}.

Finally, we point out that in the tokamak limit, when $\chi = F_0\phi$, the new reduced models derived in this section match the currently implemented JOREK tokamak models \citep{hoelzl2020the}. The $\Psi$ evolution equation, once we set $\zeta = \Omega = 0$ and drop the perpendicular components of the current, becomes
\begin{equation*}
	\left[\psi_v,\tderiv{\Psi}\right] = \left[\frac{[\Psi,\Phi]-\llderiv\Phi}{B_v},\psi_v\right] + \frac{1}{B_v}\nabla\cdot(\eta\nabla\psi_v\times\vec j^\parallel),
\end{equation*}
which can be rewritten as
\begin{equation*}
	\nabla\psi_v\cdot\nabla\times\left(\tderiv{\Psi}\nabla\chi\right) = \nabla\psi_v\cdot\nabla\times\left(\frac{\llderiv\Phi - \left[\Psi,\Phi\right]}{B_v}\nabla\chi - \eta\frac{j^\chi}{B_v^2}\nabla\chi\right).
\end{equation*}
The above equation will be satisfied as long as
\begin{equation}
	\label{eq:psipot}
	B_v\tderiv{\Psi} = \llderiv\Phi - \left[\Psi,\Phi\right] - \eta\frac{j^\chi}{B_v}
\end{equation}
is satisfied, which exactly matches the currently implemented form of the magnetic field equation in JOREK when $\chi = F_0\phi$ \citep{hoelzl2020the}. For all other equations, the fact that they match the corresponding JOREK reduced MHD equations in the tokamak limit is obvious and can be verified by simple substitution of $\chi = F_0\phi$.

\section{Energy conservation}\label{sec:encon}

We begin by showing that the new models conserve energy. If we apply the first projection operator in \eqref{eq:newprojop} to some vector field $\vec Q$, multiply the result by a test function $\Phi^*$ and integrate over the plasma volume, we can get the following expression using the identity $\nabla f\cdot\nabla\times\vec U = -\nabla\cdot(\nabla f\times\vec U)$ and integration by parts:
\begin{equation}
	\label{eq:pophi}
	\int_V \Phi^*\nabla\chi\cdot\nabla\times\left(\frac{\vec Q}{B_v^2}\right)dV = \int_V \frac{\nabla\Phi^*\times\nabla\chi}{B_v^2}\cdot\vec Q~dV,
\end{equation}
where we let $\Phi^* = 0$ on $\partial V$, so the surface integral term is zero. Doing the same with the third projection operator and test function $\zeta^*$, we have
\begin{equation}
	\label{eq:pozeta}
	\int_V \zeta^*\nabla\chi\cdot\nabla\times\left(\frac{\nabla\chi\times\vec Q}{B_v^2}\right)dV = -\int_V \frac{\nabla\zeta^*}{B_v^2}\cdot[\nabla\chi\times(\vec Q\times\nabla\chi)]dV = -\int_V \pgrad\zeta^*\cdot\vec Q~dV.
\end{equation}
We can also apply the second projection operator \eqref{eq:newprojop} to $\vec Q$, multiply by a test function $\vpar^*$, and integrate over the plasma volume:
\begin{equation}
	\label{eq:povp}
	\int_V \vpar^*\vec B\cdot\vec Q dV = \int_V \vpar^*\vec B\cdot\vec Q dV.
\end{equation}
Now, if we let $\vec Q$ be the vector momentum equation, $\vec Q \equiv \rho\partial\vec v/\partial t + \rho(\vec v\cdot\nabla)\vec v + \vec v P - \vec j\times\vec B + \nabla p = 0$, as well as $\Phi^* = \Phi$, $\vpar^* = \vpar$ and $\zeta^* = -\zeta$, and sum up the above three equations, we get
\begin{equation}
	\label{eq:vdns}
	0 = \int_V \vec v\cdot\left[\rho\tderiv{\vec v} + \rho(\vec v\cdot\nabla)\vec v + \vec v P - \vec j\times\vec B + \nabla p\right]dV,
\end{equation}
where the LHS is zero due to equations \eqref{eq:phieq}, \eqref{eq:vpeq} and \eqref{eq:zetaeq}. More importantly, if we set $\zeta = 0$ and drop equation \eqref{eq:zetaeq}, equation \eqref{eq:vdns} still holds, since with $\zeta^* = -\zeta = 0$ equation \eqref{eq:pozeta} becomes $0 = 0$, and equation \eqref{eq:zetaeq} no longer has to be satisfied in order for the LHS of equation \eqref{eq:vdns} to remain zero. Similarly, if we set $\vpar = 0$ and drop equation \eqref{eq:vpeq}, equation \eqref{eq:vdns} continues to hold due to equation \eqref{eq:povp} becoming $0 = 0$ and no longer requiring equation \eqref{eq:vpeq} to be satisfied. This property of equation \eqref{eq:vdns} being satisfied even when some terms in the velocity ansatz are missing was not present in the original models, leading to problems which we will describe in detail further in this section.

We now proceed to the magnetic field evolution equations. In the full MHD case, both of the equations \eqref{eq:mfeqns} are satisfied. These two equations are simply the vector components of Faraday's Law in the $\nabla\psi_v$ and $\nabla\chi$ directions, and as such are equivalent to the original Faraday's Law in vector form. Only two equations are needed to evolve the full magnetic field, as one degree of freedom is eliminated by the $\nabla\cdot\vec B = 0$ constraint. In the reduced MHD case, Faraday's Law is also satisfied, as one can easily see from multiplying equation \eqref{eq:psipot} by $\nabla\chi/B_v$ and taking the curl. Thus, we can take the dot product of Faraday's Law with $\vec B$ and follow the usual procedure to obtain the magnetic energy equation:
\begin{equation}
	\label{eq:magen}
	\frac{1}{2\mu_0}\tderiv{B^2} + \frac{1}{\mu_0}\nabla\cdot(\vec E\times\vec B) = -\vec v\cdot(\vec j\times\vec B) - \eta(j^*)^2,
\end{equation}
where we have $j^* = j$ in the full MHD case and $j^* = j^\parallel$ in the reduced MHD case. To complete the derivation of the energy conservation law, we rewrite equation \eqref{eq:vdns} as
\begin{equation}
	\label{eq:kineneq}
	\int_V \left[\tderiv{}\left(\frac{\rho v^2}{2}\right) + \nabla\cdot\left(\frac{\rho v^2}{2}\vec v\right)\right]dV = \int_V \left[\vec v\cdot(\vec j\times\vec B) - \vec v\cdot\nabla p - \frac{v^2}{2}P\right]dV,
\end{equation}
where we used the continuity equation. We then divide equation \eqref{eq:peq0} by $\gamma-1$ and intgrate it, together with equation \eqref{eq:magen} over the plasma volume. Adding both of the resulting equations to equation \eqref{eq:kineneq}, we obtain:
\begin{equation*}
	\begin{aligned}
		&\int_V \Bigg[\tderiv{}\left(\frac{\rho v^2}{2} + \frac{p}{\gamma-1} + \frac{B^2}{2\mu_0}\right) + \nabla\cdot\Bigg(\frac{\rho v^2}{2}\vec v + \frac{\gamma p}{\gamma-1}\vec v + \frac{\vec E\times\vec B}{\mu_0} \\
		&- \kappa_\perp\nabla_\perp T - \kappa_\parallel\nabla_\parallel T - \frac{p}{\gamma - 1}\frac{D_\perp}{\rho}\nabla_\perp\rho\Bigg)\Bigg]dV = \int_V \left(S_e - \frac{v^2}{2}P\right)dV,
	\end{aligned}
\end{equation*}
which can be rewritten as
\begin{equation}
	\frac{dE}{dt} = \oint_{\partial V} \left(\kappa_\perp\nabla_\perp T + \frac{RT}{\gamma - 1}D_\perp\nabla_\perp\rho\right)\cdot d\vec S + \int_V \left(S_e - \frac{v^2}{2}P\right)dV.
\end{equation}
Here, we have assumed that the plasma is surrounded by a perfectly conducting wall, which implies the boundary conditions $\vec v\cdot\vec n = 0$ and $\vec B\cdot\vec n = 0$, where $\vec n$ is a unit normal vector to the boundary. These conditions, along with $\Phi = \zeta = 0$ on $\partial V$, which we had assumed before to make the boundary integral terms in equations \eqref{eq:pophi} and \eqref{eq:pozeta} zero, are consistent with the boundary conditions that were used in the simulations presented in the following sections.

It is important to point out that, if we replace $P$ by $\partial\rho/\partial t + \nabla\cdot(\rho\vec v)$, exact energy conservation continues to hold even when the exact solutions are replaced by Bubnov-Galerkin finite element approximations. Indeed, a Bubnov-Galerkin solution will ensure that the expressions \eqref{eq:pophi}, \eqref{eq:pozeta} and \eqref{eq:povp} equal zero whenever the test functions $\Phi^*$, $\zeta^*$ and $\vpar^*$ are finite element basis functions. Since the finite element solutions $\Phi$, $\zeta$ and $\vpar$ are linear combinations of the basis functions, equation \eqref{eq:kineneq} will continue to hold. Similarly, since the $\Psi$ and $\Omega$ solutions will be linear combinations of the basis functions and the Bubnov-Galerkin method ensures that the weak forms of equations \eqref{eq:mfeqns} is satisfied whenever the test function is a basis function, the integral of equation \eqref{eq:magen} over volume will continue to hold. Finally, since the constant $1$ can be represented as a linear combination of basis functions, the integral of equation \eqref{eq:peq0} over volume will continue to hold, and so energy will be exactly conserved. Note, however, that this argument does not take into account temporal discretization and Fourier expansion in the toroidal direction, both of which contribute to energy conservation error in practice. However, these errors can be made small, as we discuss in section \ref{sec:examples}. The practical accuracy of energy conservation in the standard JOREK tokamak model, which is what the new model presented in section \ref{sec:changes} reduces to in the tokamak limit, is also considered in \citet{hoelzl2020the}.

The necessity of neglecting perpendicular current in the resistive term of the first equation \eqref{eq:mfeqns} when using the reduced MHD model can be clarified by deriving a magnetic field equation for the case when it is not neglected. Let $\Psi_1$ and $\Psi_2$ satisfy the equations
\begin{equation*}
	\begin{gathered}
		\left[\psi_v,\tderiv{\Psi_1}\right] = \left[\frac{[\Psi,\Phi]-\llderiv\Phi}{B_v},\psi_v\right] + \frac{1}{B_v}\nabla\cdot(\eta\nabla\psi_v\times\vec j^\parallel), \\
		\left[\psi_v,\tderiv{\Psi_2}\right] = \frac{1}{B_v}\nabla\cdot(\eta\nabla\psi_v\times\vec j^\perp),
	\end{gathered}
\end{equation*}
then $\Psi = \Psi_1 + \Psi_2$ satisfies the original first equation in \eqref{eq:mfeqns}, without the neglect of perpendicular current. The first of the two equations above is equivalent to ${\partial\vec B_1/\partial t = -\nabla\times\vec E_1}$, where $\vec B_1 = \nabla\chi + \nabla\Psi_1\times\nabla\chi$ and $\vec E_1 = -\vec v\times\vec B + \eta\vec j^\parallel$. Dotting with $\vec B = \vec B_1 + \vec B_2$ and adding $\vec B\cdot\partial\vec B_2/\partial t + \nabla\cdot(\vec E_2\times\vec B)$ to both sides, where $\vec B_2 = \nabla\Psi_2\times\nabla\chi$ and $\vec E_2 = \eta\vec j^\perp$, we get
\begin{equation*}
	\frac{1}{2}\tderiv{B^2} + \nabla\cdot(\vec E\times\vec B) = \mu_0[(\vec v\times\vec B)\cdot\vec j - \eta j^2] + \vec B\cdot\tderiv{\vec B_2} + \vec B\cdot\nabla\times\vec E_2.
\end{equation*}
The evolution equation for $\Psi_2$ can be written as $(\partial\vec B_2/\partial t)^{\psi_v} = -(\nabla\times\vec E_2)^{\psi_v}$, and so the last two terms in the RHS of the above equation will only cancel partially, leaving behind the nonconservative terms $B_{\beta_v}\partial B_2^{\beta_v}/\partial t + B_{\beta_v}(\nabla\times\vec E_2)^{\beta_v} + (\nabla\times\vec E_2)^\chi$.

Finally, we show why the projection operators used in our original model allowed for nonphysical generation of kinetic energy. To begin with, the velocity field can be written in a covariant form:
\begin{equation}
	\label{eq:covvel}
	\vec v = B_v^2\pgrad\widetilde{\Phi}\times\vec e_\chi + \widetilde{\vpar}\nabla\chi + B_v^2\vec e_\chi\times(\nabla\widetilde{\zeta}\times\nabla\chi),
\end{equation}
as opposed to the contravariant form \eqref{eq:anz}. The fact that any vector field can be written in this way is easy to show using the same arguments as in \citet{nikulsin2019a}, except using the projection operators \eqref{eq:newprojop} instead of the operators \eqref{eq:oldprojop}. Note that the projection operators \eqref{eq:newprojop} are orthogonal to all but one term in the covariant velocity ansatz \eqref{eq:covvel}:
\begin{equation*}
	\begin{gathered}
		\nabla\chi\cdot\nabla\times\left(\frac{1}{B_v^2}\vec v\right) = -\nabla\cdot(\pgrad\widetilde{\Phi}), \\
		\vec B\cdot\vec v = B^\chi\widetilde{\vpar}, \\
		\nabla\chi\cdot\nabla\times\left(\frac{1}{B_v^2}\nabla\chi\times\vec v\right) = \nabla\cdot(B_v^2\pgrad\widetilde{\zeta}).
	\end{gathered}
\end{equation*}

Now, applying the first projection operator in \eqref{eq:oldprojop} to a vector field $\vec Q$ and then multiplying by a test function $\Phi^*$ and integrating, we get:
\begin{equation}
	\int_V \Phi^*\nabla\chi\cdot\nabla\times[\nabla\chi\times(\vec e_\chi\times\vec Q)]dV = -\int_V B_v^2(\pgrad\Phi^*\times\vec e_\chi)\cdot\vec QdV,
\end{equation}
where we again used integration by parts to get the RHS. Similarly, for the third projection operator in \eqref{eq:oldprojop}, we can write:
\begin{equation}
	\label{eq:3opo}
	\int_V \zeta^*\nabla\cdot[B_v^2\nabla\chi\times(\vec e_\chi\times\vec Q)]dV = \int_V B_v^2[\vec e_\chi\times(\nabla\zeta^*\times\nabla\chi)]\cdot\vec QdV,
\end{equation}
and for the second projection operator, we simply multiply by $\vpar^*$ and integrate, without doing any transformations:
\begin{equation}
	\int_V \vpar^*\nabla\chi\cdot\vec QdV = \int_V \vpar^*\nabla\chi\cdot\vec QdV.
\end{equation}
In order to get $\int_V \rho\vec v\cdot\vec QdV = 0$ when we set $\vec Q \equiv \partial\vec v/\partial t + (\vec v\cdot\nabla)\vec v + \vec v P/\rho - \vec j\times\vec B/\rho + \nabla p/\rho = 0$, we need $\Phi^* = -\widetilde{\Phi}$, $\vpar^* = \widetilde{\vpar}$, $\zeta^* = \widetilde{\zeta}$ and $\rho = \const$. The first three conditions pose no problem in the full MHD case, but if we try to set $\zeta = 0$, it does not imply $\widetilde{\zeta} = 0$, and thus when we drop the evolution equation for $\zeta$, equation \eqref{eq:3opo} is no longer satisfied, and so the kinetic energy equation \eqref{eq:kineneq} cannot be satisfied. In addition, because the vector momentum equation was divided by density before the projection operators were applied, unless the density is held constant, the kinetic energy equation \eqref{eq:kineneq} also cannot be satisfied even when $\zeta\neq 0$. Thus, both the reduction and spatially varying density can lead to nonphysical generation of kinetic energy.

\section{Numerical examples}\label{sec:examples}

\begin{figure}
	\includegraphics[scale=0.8]{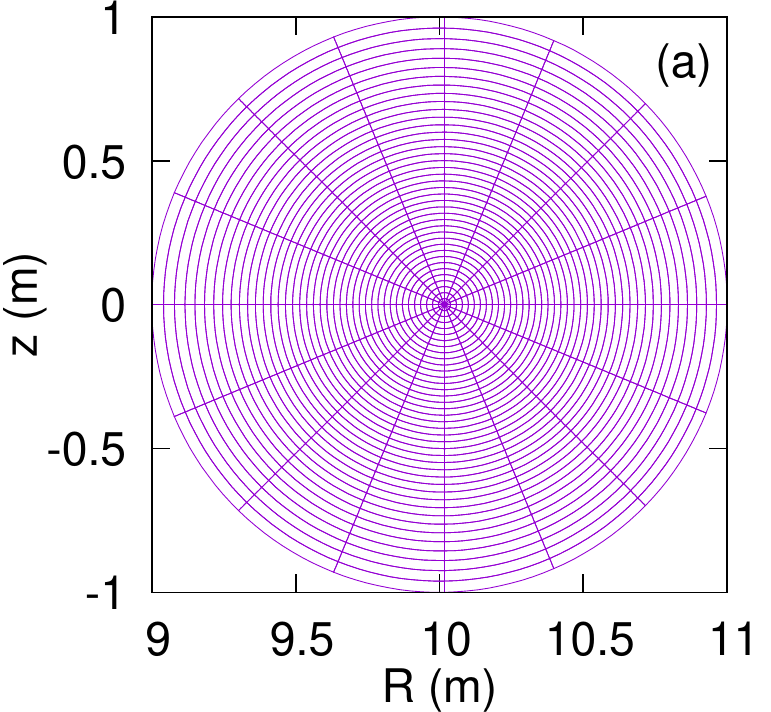}\hfill
	\raisebox{0.5cm}{\includegraphics[scale=0.25]{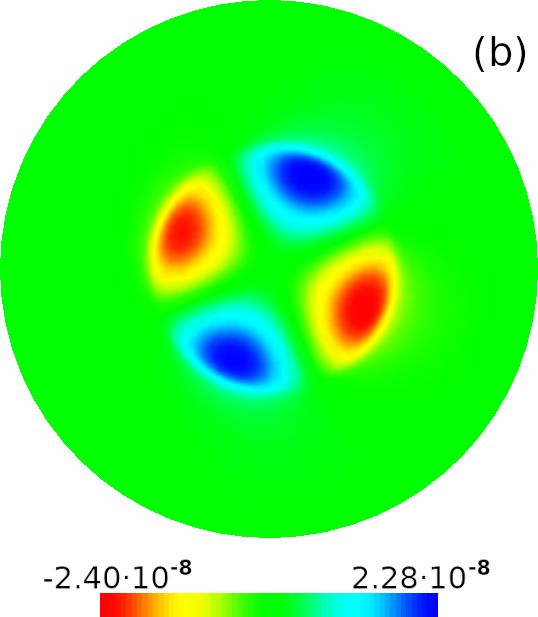}}
	\caption{A flux-aligned grid used for simulating the tearing mode (a), and the $n=1$ Fourier mode of $\psi = F_0\Psi$ (JOREK units) in the standard tokamak model at $t=50000$ Alfv\'en times (b).}
	\label{fig:tearing}
\end{figure}

In this section, we consider several simulations with the JOREK code for a test case of a tearing mode in a circular cross section tokamak with an aspect ratio of 10 (Figure \ref{fig:tearing}). We compare the reduced model without parallel flow from the original set of stellarator-capable models, as presented in \citet{nikulsin2019a}, to the standard JOREK tokamak model without parallel flow \citep{hoelzl2020the}. Note that in the tokamak limit $\chi = F_0\phi$, the latter is equivalent to the the model without parallel flow from the set of new stellarator-capable models, as introduced in section \ref{sec:changes}. In addition, we point out that the standard tokamak reduced MHD model in JOREK is known to accurately reproduce tearing modes when compared to full MHD \citep{Pamela2020,Haverkort2016}, thus comparing to the standard tokamak model is sufficient.

In the test cases considered below, $F_0 = 10~\mathrm{T\cdot m}$. The initial conditions were set up by solving the Grad-Shafranov equation with $FF'(\psi_n) = 1.173(1-\psi_n)$ in units of $\mathrm{T^2\cdot m}$, where $\psi_n = (\psi - \psi_{axis})/(\psi_{edge}-\psi_{axis})$, and $p(\psi_n)=0$, except for the last case. In the tokamak limit, the $\psi$ in the Grad-Shafranov equation is related to the $\Psi$ in the magnetic field ansatz \eqref{eq:anz} by $\psi = F_0\Psi$. The density was initialized to be constant at $3.346\cdot 10^{-7}\mathrm{kg/m^3}$, corresponding to $10^{20}$ deuterium ions per cubic meter, and $\Phi$ was initialized to zero. A viscosity of $\mu = 5.159\cdot 10^{-8}~\mathrm{kg/(m\cdot s)}$ was used in all simulations, which corresponds to $10^{-5}$ in JOREK units for $\mu_\perp^t$ in the standard tokamak model.\footnote{The viscosity $\mu_\perp^t$ in the standard tokamak model is not the same as the $\mu_\perp$ of equation \eqref{eq:phieq}, instead we have $\mu_\perp^t = F_0^2\mu_\perp$.} When using the original stellarator model, the kinematic viscosity was set to be $\nu = \mu/\rho_0 = 0.1542~\mathrm{m^2/s}$ ($10^{-7}$ in JOREK units). Finally, since JOREK discretizes the toroidal direction by approximating all solutions as toroidal Fourier series, in all simulations shown in this section, the Fourier series are truncated after the first mode for the sake of simplicity, keeping only the axisymmetric ($n=0$) and the $n=1$ terms, unless noted otherwise. Finite elements are used for discretization in the poloidal plane. Temporal discretization is done via the Crank-Nicolson scheme, i.e. an average of the forward and backward Euler schemes. Given equations of the type
\begin{equation*}
    \sum_i a_i(\vec U)\tderiv{A_i(\vec U)} = B(\vec U),
\end{equation*}
where $\vec U$ is the set of all unknowns in the model and $a_i$, $A_i$ and $B$ are the expressions that comprise the equations and can involve spatial derivatives. The Crank-Nicolson scheme gives:
\begin{equation*}
    \sum_i \left[\frac{1}{2}a_i(\vec U~^{n+1}) + \frac{1}{2}a_i(\vec U~^n)\right][A_i(\vec U~^{n+1}) - A_i(\vec U~^n)] = \Delta t\left[\frac{1}{2}B(\vec U~^{n+1}) + \frac{1}{2}B(\vec U~^n)\right],
\end{equation*}
where $n$ in the superscript refers to the already calculated values of the model variables in the current time step, and $n+1$ refers to the next time step, yet to be calculated. In addition, a temporal linearization is done around the current time step, leading to the actual numerical scheme that is implemented in the code:
\begin{equation*}
    \sum_i a_i(\vec U~^n)\left.\frac{\partial A_i}{\partial\vec U}\right|_{\vec U^n}\cdot\delta\vec U~^n = \Delta t\left[B(\vec U~^n) + \frac{1}{2}\left.\frac{\partial B}{\partial\vec U}\right|_{\vec U^n}\cdot\delta\vec U~^n\right],
\end{equation*}
where $\delta\vec U~^n = \vec U~^{n+1} - \vec U~^n$. For more details on the numerical implementation, see \citet{czarny2008bezier,huysmans2007mhd,hoelzl2020the}.

\begin{figure}
	\includegraphics[scale=0.55]{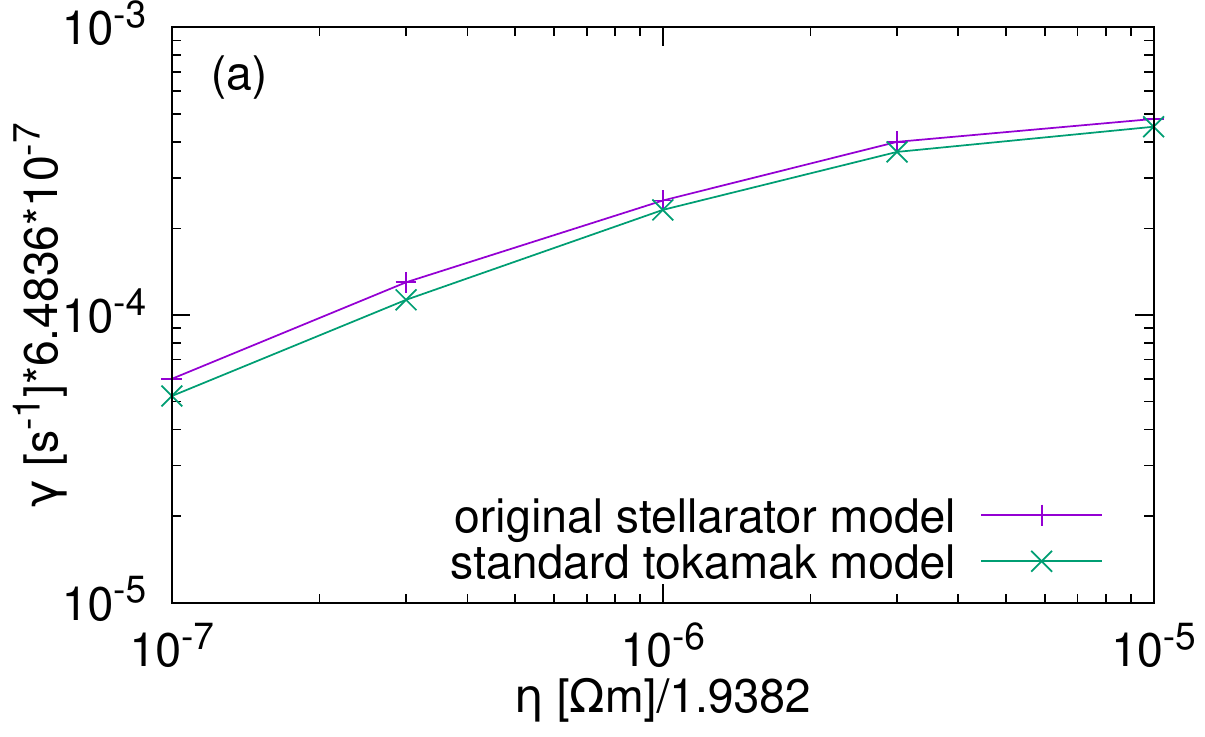}\hfill\includegraphics[scale=0.55]{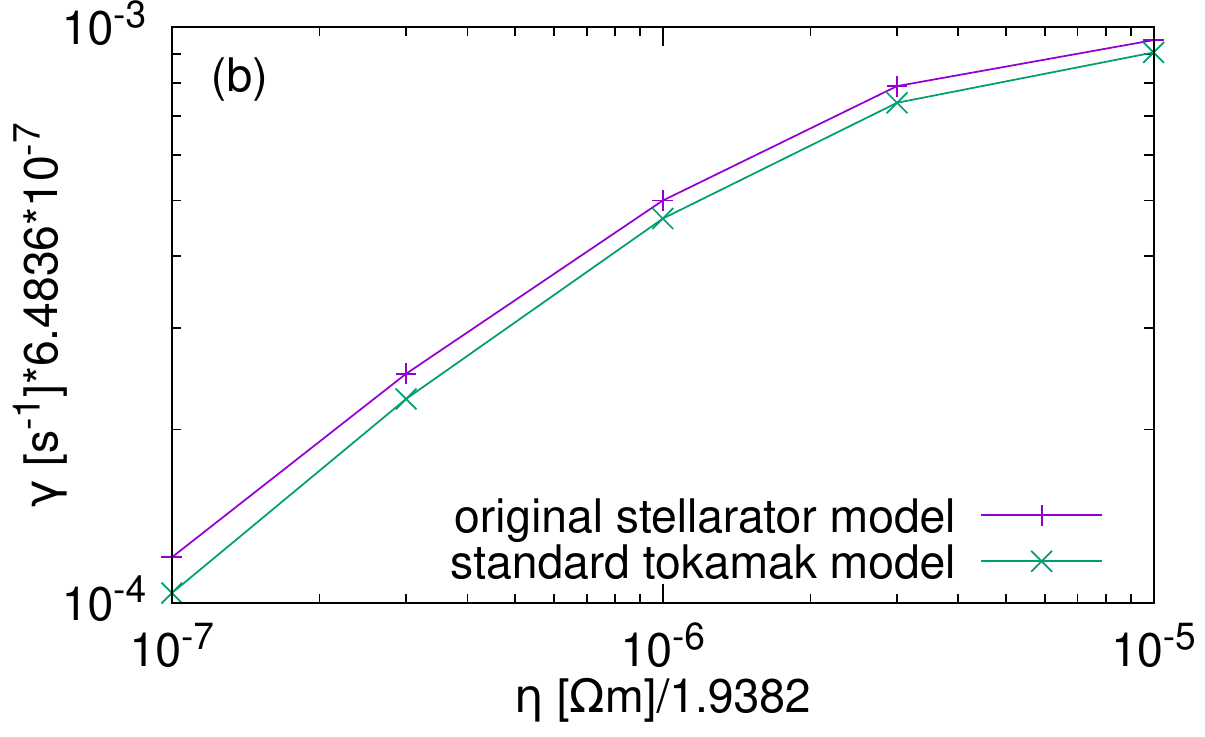}
	\caption{Tearing mode growth rates at various plasma resistivities in the original stellarator model and the standard tokamak model. Plot (a) shows the growth rates for the $n=1$ toroidal mode in simulations with $n=0,1$ and (b) shows the growth rates for the $n=2$ toroidal mode in simulations with $n=0,...,4$. The growth rates of the $n=2$ modes are double those of the $n=1$ modes, indicating that the $n=2$ modes are not naturally unstable but excited by non-linear mode coupling.}
	\label{fig:grates}
\end{figure}

To begin with, we tested the original stellarator model \citep{nikulsin2019a} in the linear regime by calculating the tearing mode growth rates at different resistivities and comparing them to the growth rates obtained from the standard tokamak model \citep{hoelzl2020the}. In both cases, we did a spatial and temporal resolution scan at each resistivity to ensure that the growth rate that we recorded was converged. We then repeated the simulations with Fourier modes up to and including $n=4$ with the same spatial and temporal resolutions at which the growth rates converged for the $n=0,1$ simulations. The results of the $n=0,1$ simulations are shown in Figure \ref{fig:grates}~a. Figure \ref{fig:grates}~b shows the growth rates of the nonlinearly driven $n=2$ modes in the simulations with $n=0,...,4$; these modes are not inherently unstable, but are destabilized by the unstable $n=1$ mode. As can be seen, both models show decent agreement. At each resistivity, the $n=2$ growth rates are roughly twice the $n=1$ growth rates, as expected. In both models at higher resistivities, the $n=3$ and $n=4$ modes are destabilized shortly before the onset of the nonlinear regime, and so the corresponding growth rates (not shown here) do not plateau, but rather peak at values roughly three and four times the value of the $n=1$ growth rate and then decline rapidly as nonlinear saturation is reached. All of the growth rates shown in Figure \ref{fig:grates} were calculated for $\beta=0$, and the pressure was not evolved. In the case of the the original stellarator model, this amounts to not using the full energy conservation equation (third equation in \eqref{eq:vrmhd}). For clarity, we explicitly list the equations solved in Table \ref{tab:eqs_solved}.

\begin{table}
    \centering
    \begin{tabular}{c|c}
        Original stellarator model & Standard tokamak model \\
        \hline
        $\tderiv{\rho} = - B_v\left[\frac{\rho}{B_v^2},\Phi\right] + P$ & $\tderiv{\rho} =  \frac{1}{R}\left[R^2\rho,u\right] + P$ \\
        $\left[\psi_v,\tderiv{\Psi}\right] = \left[\frac{[\Psi,\Phi]-\llderiv\Phi}{B_v},\psi_v\right] + \frac{1}{B_v}\nabla\cdot(\eta\nabla\psi_v\times\vec j_d^\parallel)$ & $\tderiv{\psi} = -F_0\frac{\partial u}{\partial\phi} + R\left[\psi,u\right] - \eta R^2 j_d^\phi$ \\
        $\pLap\tderiv{\Phi} = \nabla\cdot\Big[\frac{B_v\llderiv v^2}{2}\pgrad\Psi + \omega_\chi\nabla\Phi\times\nabla\chi$ & $\nabla\cdot\left(R^2\rho\pgrad\tderiv{u}\right) = \frac{-1}{2R}\left[R^2\rho,v^2\right]$ \\
        $- v_\chi B_v^2\vec\omega^\perp + \frac{B^2}{\rho}\vec j^\perp - \frac{j_\chi B_v^2}{\rho}\nabla\Psi\times\nabla\chi - \frac{1}{\rho}\vec f_b\times\nabla\chi$ & $+ \frac{1}{R}\left[R^4\rho\omega^\phi,u\right] + \frac{F_0}{R^2}\frac{\partial}{\partial\phi}(R^2 j^\phi) + \frac{1}{R}\left[R^2 j^\phi,\psi\right]$ \\
        $- \frac{P}{\rho}\pgrad\Phi\Big] + \nu\Delta\pLap\Phi - \nu_h\Delta^2\Delta^\perp\Phi$ & $+ \nabla\cdot(\mu_\perp^t\pgrad\pLap u)$ \\
        $\vec B = \nabla\chi + \nabla\Psi\times\nabla\chi; \qquad \vec v = \frac{\nabla\Phi\times\nabla\chi}{B_v^2}$ & $\vec B = F_0\nabla\phi + \nabla\psi\times\nabla\phi; \qquad \vec v = R^2 \nabla\phi\times\nabla u$
    \end{tabular}
    \caption{The zero-$\beta$ forms of the two reduced MHD models that are compared in this section, namely the original stellarator model from \hbox{\citet{nikulsin2019a}} and the standard tokamak model without field-aligned flow from \hbox{\citet{franck2015energy}}. Note that the new set of equations presented in section \ref{sec:changes} reduces to the standard tokamak model when $\zeta = \vpar = \Omega = 0$ and $\chi = F_0\phi$, except for the term containing $P$ (eighth term in the RHS of equation \eqref{eq:phieq}). In the cases considered, $P = \nabla\cdot(D_\perp\nabla_\perp\rho)$. When using the original stellarator model, we set $\chi = F_0\phi$ and $\psi_v = R$, where $R$ is the distance from the central axis of symmetry; in addition a subscript $\chi$ means a dot product of the corresponding vector with $\vec e_\chi = \vec B/B_v^2$, and $\vec f_b = -FF'\nabla\Psi|_{t=0}/R^2$ is the force balancing term (see Appendix \ref{sec:appendix}). In order for the initial condition to be a true equilibrium, we have introduced $\vec j_d = \vec j - \vec j_0$, where $\vec j_0$ is the current at $t=0$.}
    \label{tab:eqs_solved}
\end{table}

\begin{figure}
	\includegraphics[scale=0.55]{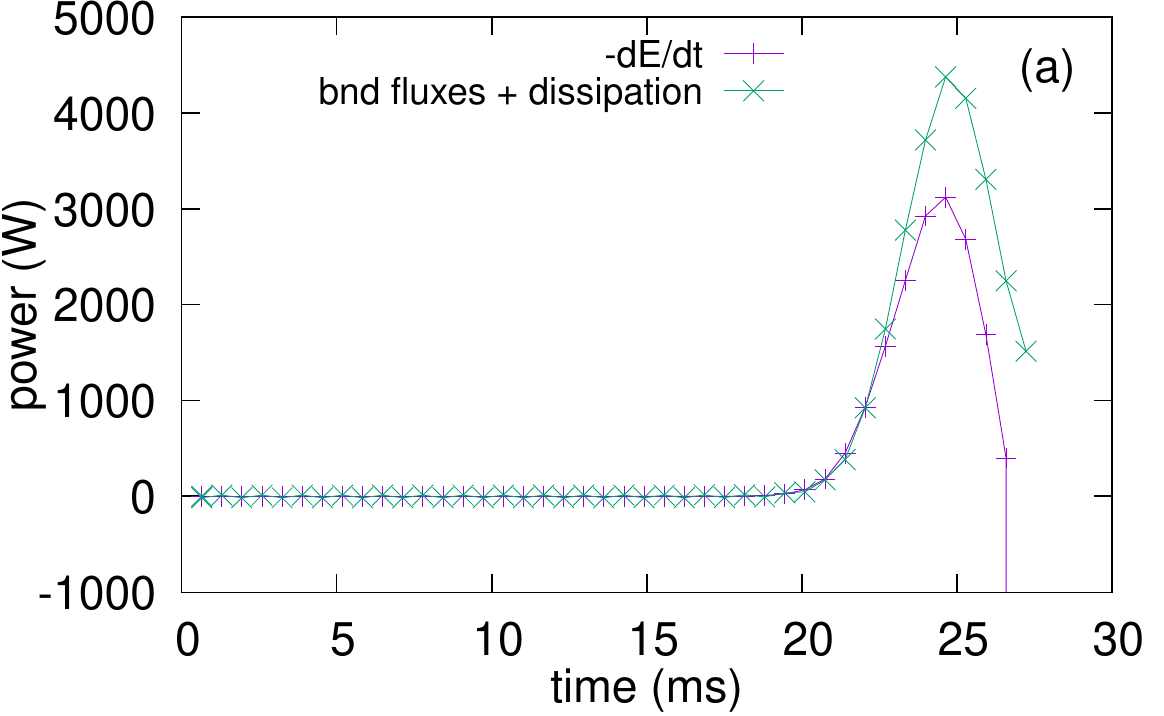}\hfill\includegraphics[scale=0.55]{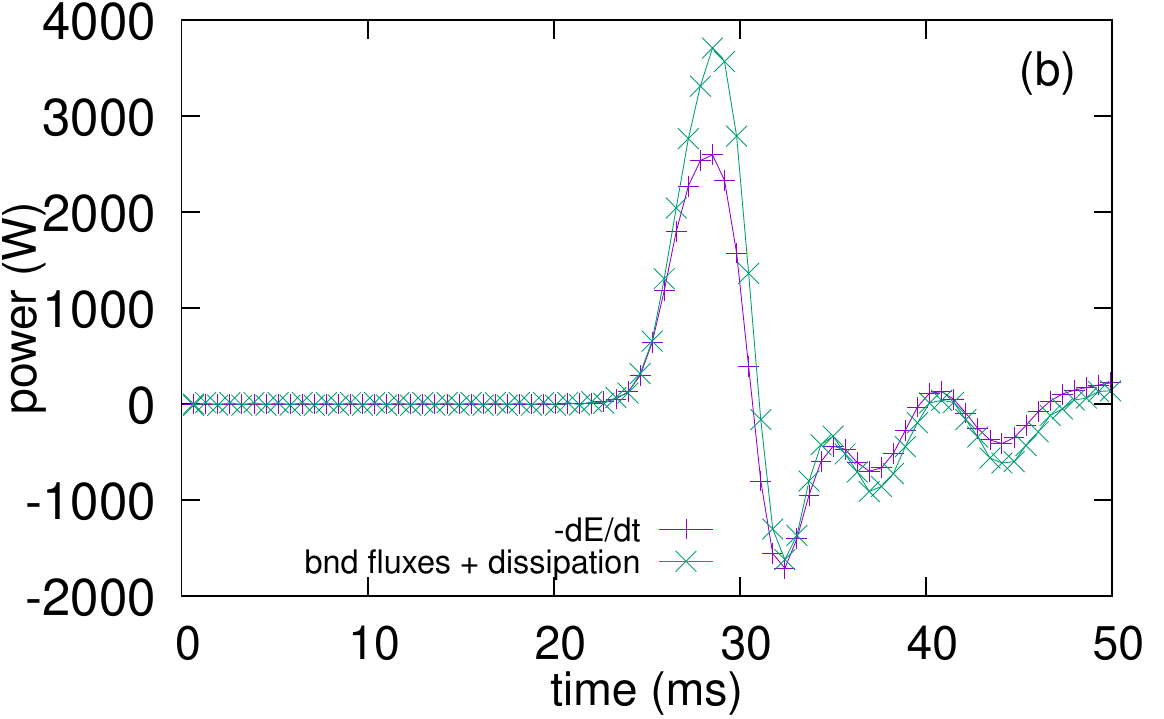} \\
	\includegraphics[scale=0.55]{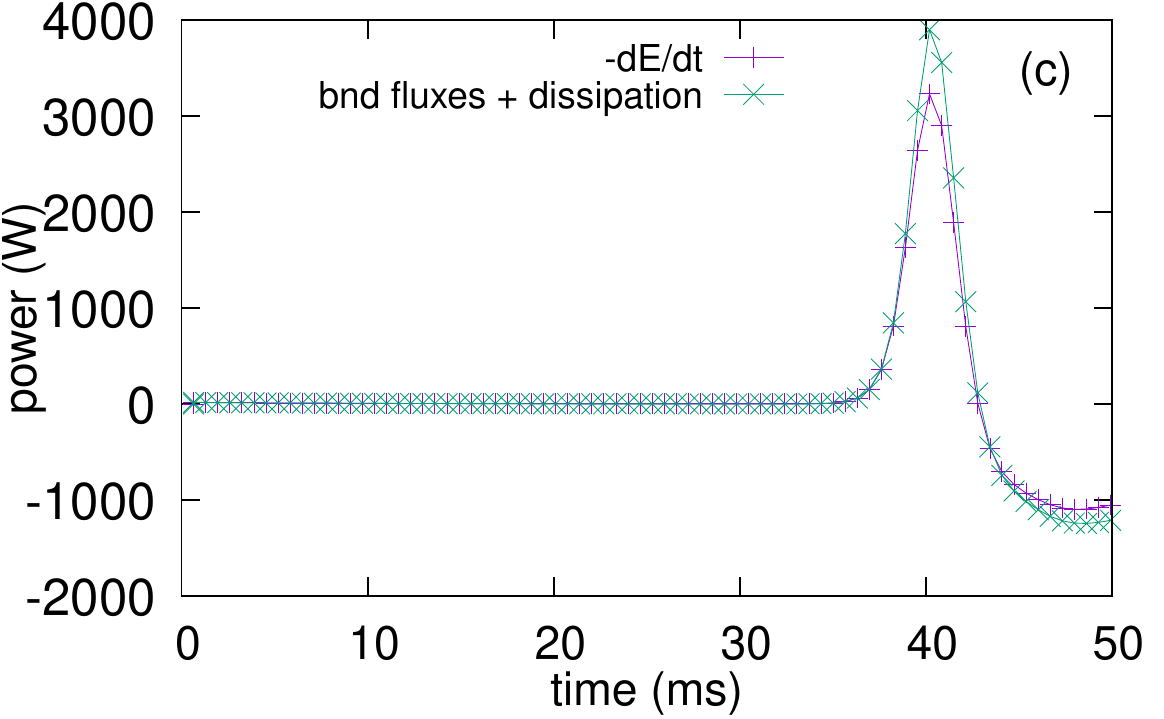}\hfill\includegraphics[scale=0.55]{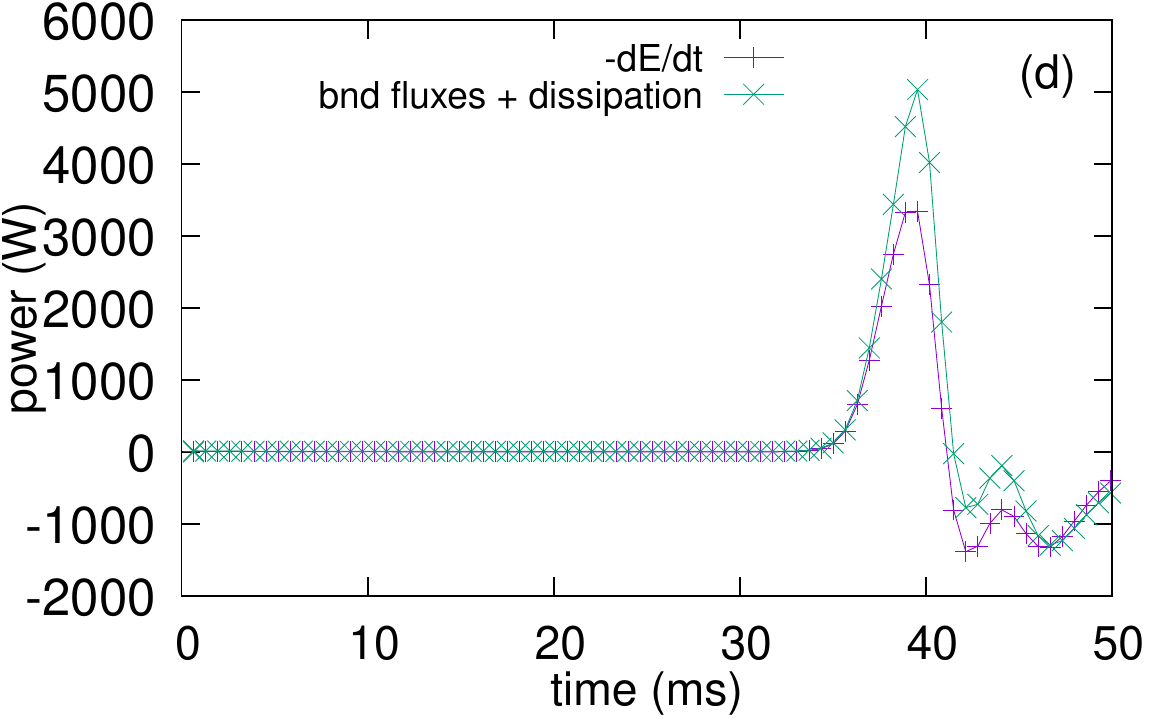}
	\caption{The negative rate of change of total energy compared to the physical energy loss rate in (a) the original stellarator model without artificial dissipation, (b) the original stellarator model with artificial dissipation, (c) the standard tokamak model/new stellarator model, and (d) the original stellarator model with equation \eqref{eq:psipot} replacing the $\Psi$ equation in Table \ref{tab:eqs_solved}.}
	\label{fig:encon}
\end{figure}

\begin{figure}
    \includegraphics[scale=0.55]{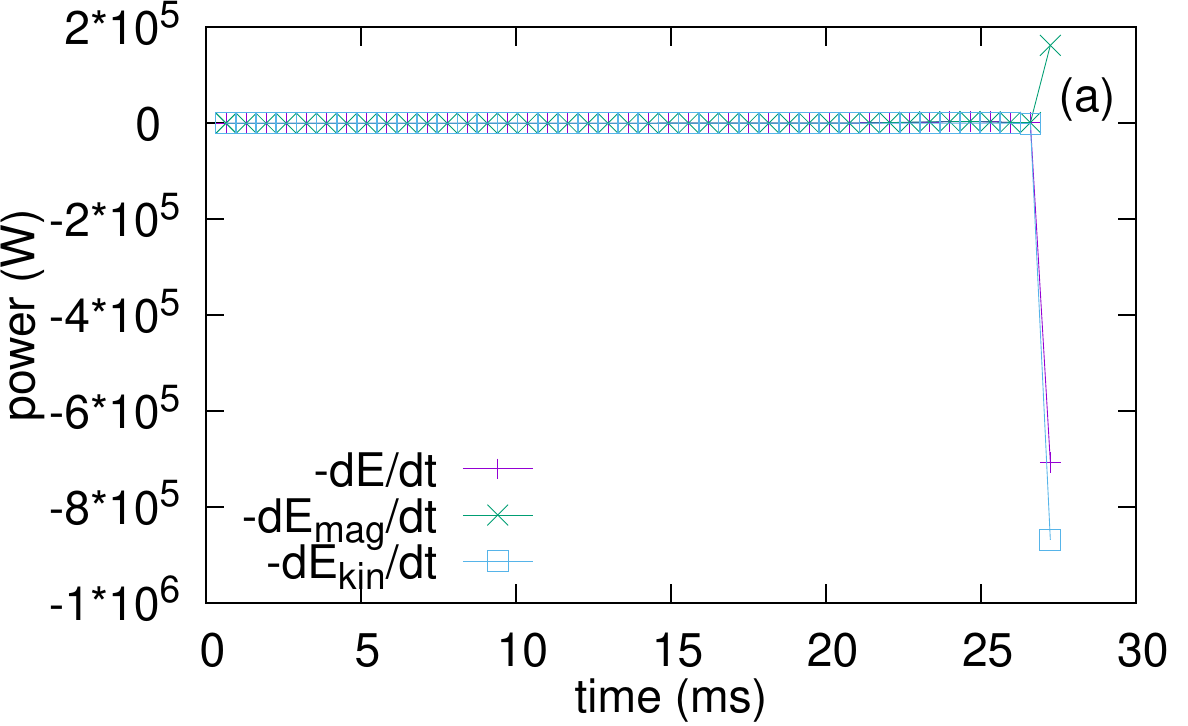}\hfill\includegraphics[scale=0.55]{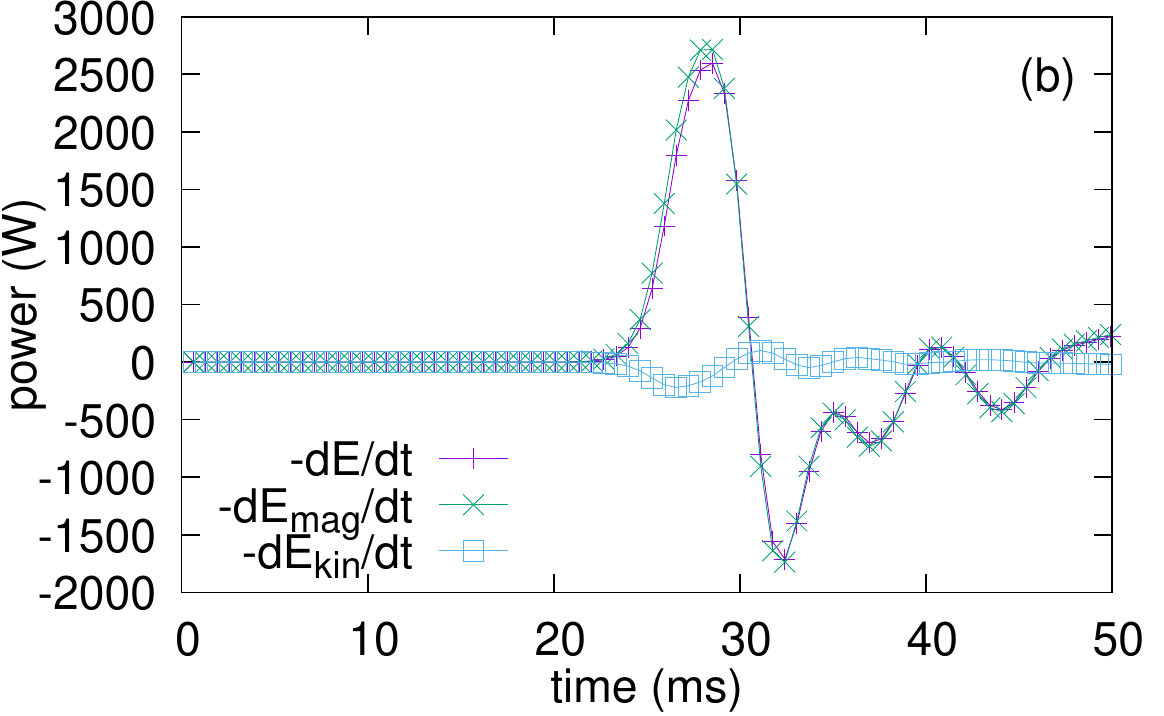}
    \caption{The negative rate of change of total energy $-dE/dt$ from Figure \ref{fig:encon} plotted alongside the negative rates of change of the magnetic and kinetic energies, where $dE/dt = dE_{mag}/dt + dE_{kin}/dt$. Plot (a) shows the negative rates of change in the original stellarator model without artificial dissipation, and (b) shows the same in the case with artificial dissipation.}
    \label{fig:dEdt}
\end{figure}

While the original stellarator model \citep{nikulsin2019a} performs decently in the linear regime and at $\beta=0$, two problems can arise after nonlinear saturation. First, as discussed in the previous section, nonphysical kinetic energy can be generated. The rate at which it is generated is negligible in the linear regime, but can become significant as saturation is reached. This can be seen in Figure \ref{fig:encon}, where we consider the tearing mode simulation with a resistivity of $10^{-5}$ JOREK units ($1.9382\cdot 10^{-5}\mathrm{\Omega\cdot m}$). In Figure \ref{fig:encon}, $-dE/dt$ is plotted and compared to the physical energy loss rate. We define $E = \int_V\mathcal{E}dV$, where $\mathcal{E}$ is given in \eqref{eq:vrmhd}; this is the integrated total energy of the system at a given time. The physical energy loss rate is defined as the sum of the energy fluxes across the plasma boundary and the volume integral of energy sinks due to resistive and viscous dissipation (conversion to internal energy is not accounted for since pressure is not evolved, so dissipated energy is lost). If the inward energy fluxes are greater than outward fluxes plus sinks, the physical loss rate is negative, i.e. energy is gained. As can be seen in Figure \ref{fig:encon}~a, the difference between $-dE/dt$ and the physical loss rate is negligible in the linear regime (until approximately 20~ms) and grows rapidly with the onset of saturation, due to the uncontrollable increase in kinetic energy (Figure \ref{fig:dEdt}~a). The simulation is stopped at $\sim$27~ms, as it would crash shortly thereafter if allowed to continue. In order to allow the simulation to continue after saturation, we can introduce an artificial dissipation. Figure \ref{fig:encon}~b shows $-dE/dt$ and physical energy loss rate for a simulation with artificial dissipation in the form of a hyperviscosity term (with $\nu_h = 1.08\cdot 10^{-3}~\mathrm{m^4/s}$, corresponding to $7\cdot 10^{-10}$ in JOREK units) in the evolution equation for $\Phi$. In this case, the physical energy loss rate does not include the hyperviscous dissipation. As can be seen, with artificial dissipation, the energy behaves much more reasonably, though it is still not conserved, hence the mismatch between the total rate of energy change and physical energy loss. The reason for such good behavior is that the non-conservation comes mostly from the kinetic energy, which is kept from exploding by the hyperviscosity term. In addition, the hyperviscosity prevents the plasma from responding too quickly to numerical errors in the magnetic field, thus stabilizing it against numerical instabilities that arise from the $\Psi$ evolution equation (discussed below).

One can introduce a finite pressure into the original stellarator model by using equation \eqref{eq:peq} without affecting the energy conservation by much, since all of the error comes from the kinetic energy, which is usually not dominant in fusion-relevant situations. However, we find that the original stellarator model cannot correctly reproduce the growth rates when $\beta$ becomes non-negligible, likely because the pressure terms in the original stellarator model (\citet{nikulsin2019a}, not shown in Table \ref{tab:eqs_solved}) contain either $\nabla\rho$ or $\llderiv p$, both of which are zero initially, and so they make the pressure terms small compared to the Lorentz force terms.

Finally, for the sake of comparison, we show the same kind of graph for a simulation with the standard tokamak model/new stellarator model without parallel flow (Figure \ref{fig:encon}~c), both of which are equivalent in the tokamak limit. Note the small discrepancy between $-dE/dt$ and the physical energy loss rate in the last plot; unlike the discrepancy in Figure \ref{fig:encon}~b, this discrepancy is purely numerical. As the arguments of section \ref{sec:encon} continue to apply even after the exact solutions are replaced by finite element approximations, the discrepancy is not due to poloidal discretization, and is independent of the resolution of the finite elements, which we have confirmed by running simulations with higher resolution. Instead, the two sources of error are: a) the temporal discretization, where a small enough time step is used in practice to diminish the deviation, and b) the truncation of the toroidal Fourier series, where a sufficiently large toroidal resolution is used in practice to avoid this effect. We have carried out further simulations (not shown here), where we have confirmed that by decreasing the time step size and increasing the number of Fourier modes kept, one can make the discrepancy arbitrarily small.

\begin{figure}
    \stackinset{c}{0.25cm}{b}{1.2cm}{\rotatebox{-20}{\rule{1.08cm}{0.8pt}}}{
	\stackinset{c}{1.87cm}{b}{1.25cm}{\rotatebox{90}{\rule{0.35cm}{0.8pt}}}{\includegraphics[scale=0.55]{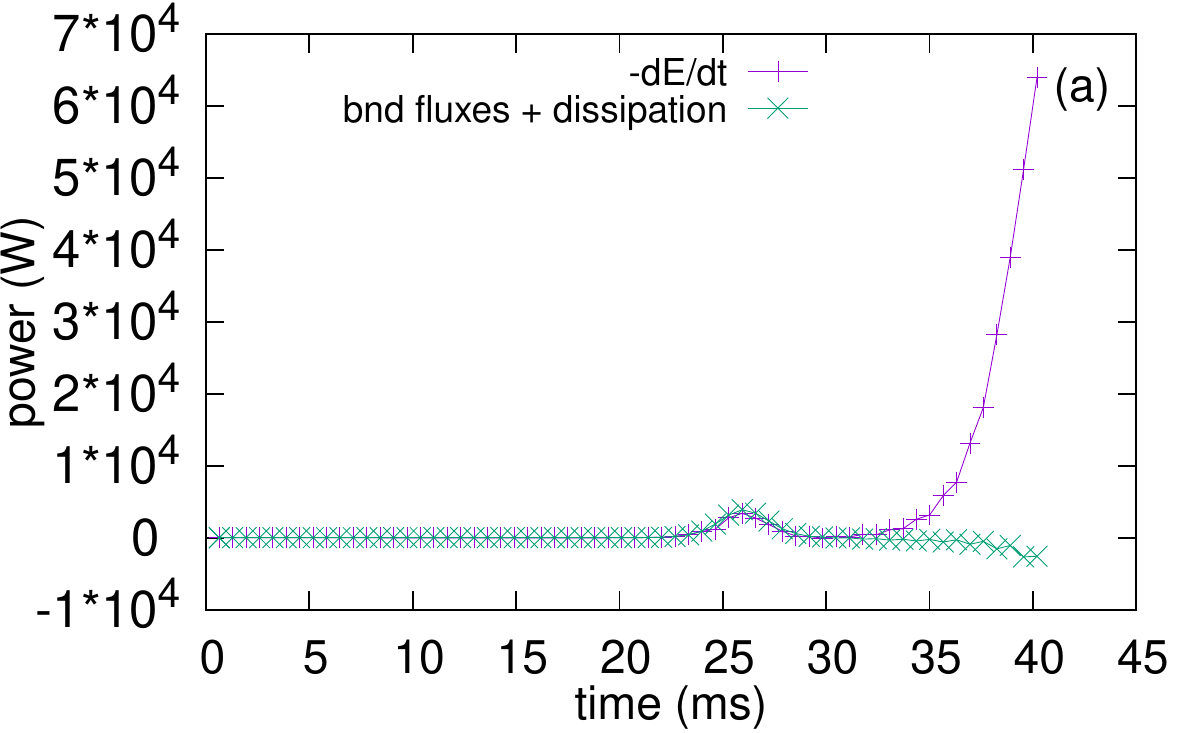}}}
    \hspace{-4.35cm}\raisebox{1.6cm}{\includegraphics[scale=0.23]{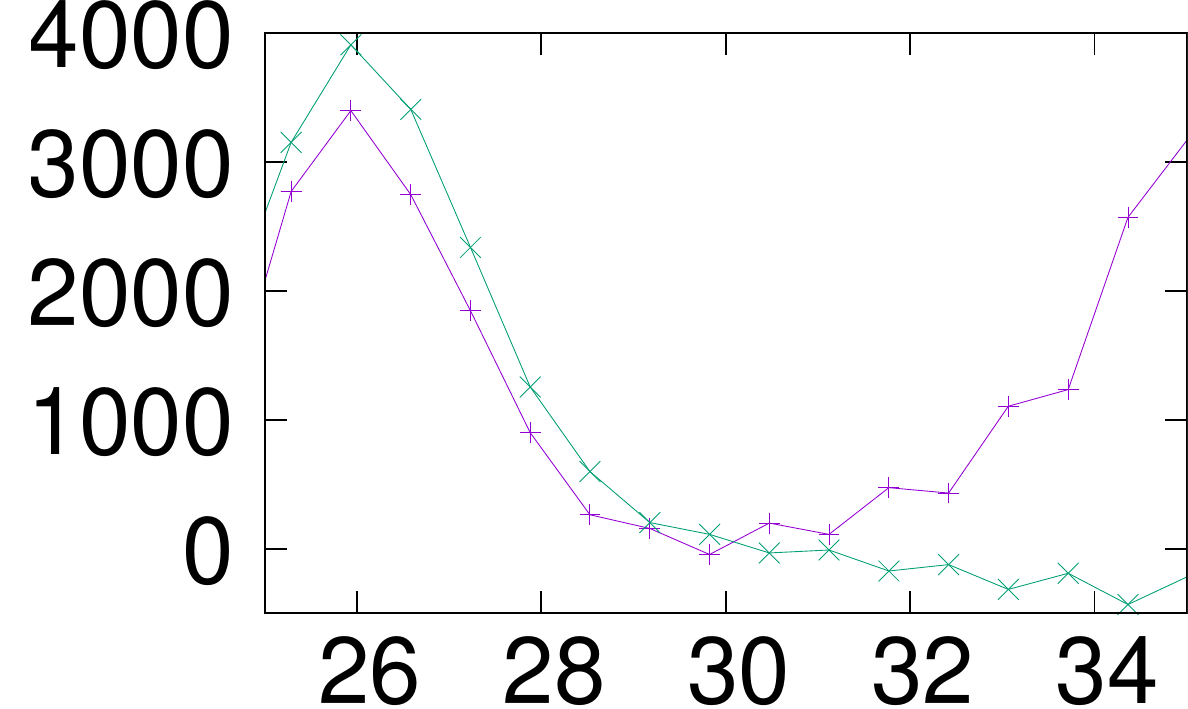}}
    \hfill\includegraphics[scale=0.55]{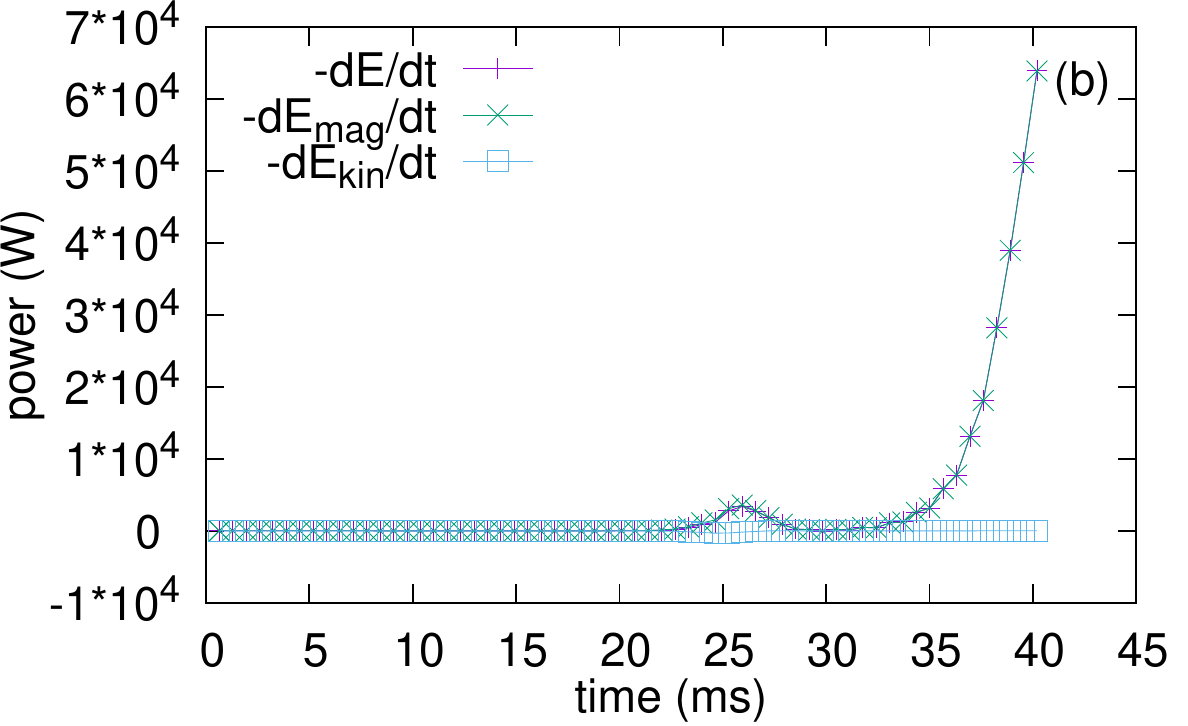}
    \caption{A simulation of a tearing mode with the standard tokamak model when the $\psi$ evolution equation is replaced by the corresponding equation from the original stellarator model. Plot (a) is a comparison of the negative rate of change of total energy $-dE/dt$ to the physical loss rate and (b) shows the negative rates of change of magnetic and kinetic energy alongside $-dE/dt$. The inset in (a) zooms in on times between 25~ms and 35~ms.}
    \label{fig:oldpsi_newphi}
\end{figure}

The second problem is that, while looking benign, the $\Psi$ evolution equation can often produce ill-conditioned time stepping matrices, resulting in numerical instabilities. In fact, the $\Psi$ equation is what destabilizes the kinetic energy, causing it to explode in Figure \ref{fig:dEdt}~a. If we replace the $\Psi$ evolution equation in the original stellarator model (see Table \ref{tab:eqs_solved}) by equation \eqref{eq:psipot}, we can run the simulations well into the post-saturation regime without needing to introduce hyperviscosity, as can be seen in Figure \ref{fig:encon}~d, albeit with a higher energy conservation error than when hyperviscosity is present. On the other hand, we can replace the $\psi$ evolution equation in the standard tokamak model by the corresponding equation from the original stellarator model (Table \ref{tab:eqs_solved}), thus testing the $\Psi$ equation from the original stellarator model in a setting where energy is conserved in the continuous limit. In this case (Figure \ref{fig:oldpsi_newphi}~a), energy conservation issues begin around the same time that the original model without hyperviscosity would have crashed (Figure \ref{fig:encon}~a). However, instead of exploding, the energy rapidly decreases, and so this simulation can run a bit longer before crashing. Figure \ref{fig:oldpsi_newphi}~b shows that the kinetic energy plays no part in this numerical instability; all of the energy conservation error comes from the buildup of numerical errors in the magnetic energy. As can be seen, while the $\Psi$ evolution equation in the original stellarator model in Table \ref{tab:eqs_solved} is equivalent to equation \eqref{eq:psipot} in the analytical sense, numerically it behaves very differently. Finally, we verified that replacing the psi evolution equation in all models considered in this section does not have any significant impact on the growth rates.

To conclude this section, we note that we had initially intended to use the full energy conservation equation, as shown in \eqref{eq:vrmhd}, to evolve pressure and ensure energy conservation. However, using the full energy conservation equation would make the internal energy into a reservoir, drawing energy from it when there is a nonphysical gain of kinetic energy, or depositing energy into it if there is a nonphysical loss of kinetic energy. This shifts the energy conservation error to pressure while ensuring total energy conservation. While this may be acceptable when the internal energy is much larger than the kinetic energy, using the full energy conservation equation would thus disallow low $\beta$ simulations. It is much better to use equation \eqref{eq:peq} with the original stellarator model instead of the full energy conservation equation, although, as noted above, the original stellarator model is unable to accurately reproduce growth rates when $\beta$ is not negligible.

\section{Approximate conservation of momentum}\label{sec:momentum}

Reduced MHD models usually do not conserve linear momentum exactly, except for some special cases. Locally, this can be seen from the fact that there are only two velocity variables, $\Phi$ and $\vpar$, in reduced MHD, so the three components of the full MHD momentum equation cannot be satisfied simultaneously. However, different reduced MHD models can have different amounts of error in the momentum conservation. In the remainder of this section, we will first compare momentum conservation properties of the original model presented in \citet{nikulsin2019a} to those of the new model derived in section \ref{sec:changes}, and then show how global momentum behaves in the new model using some numerical examples. When discussing momentum conservation, we do not exclude exchange of momentum with the walls, such as in the case of a vertical displacement event. Exchange of momentum is a physical process, and, while the total momentum of the simulated system can change, no momentum is created or destroyed. In this section, we are concerned with the nonphysical generation of momentum within the plasma due to approximations made in reduced MHD.

\subsection{General local momentum conservation properties of the reduced models}

We begin by reviewing the momentum conservation properties of the original reduced model, which were discussed in detail in \citet{nikulsin2019a}. The action of the first projection operator \eqref{eq:oldprojop} on a vector $\vec Q$ can be written as $\nabla\chi\cdot\nabla\times[\nabla\chi\times(\vec e_\chi\times\vec Q)] = \nabla\chi\cdot\nabla\times(\vec e_\chi Q^\chi - \vec Q)$. Thus, if $\vec Q$ is the full MHD momentum equation \eqref{eq:vrmhd}, the action of the projection operator corresponds to dropping the two vector components of a vorticity-like equation, $\nabla\times(\vec e_\chi Q^\chi - \vec Q)$, that are perpendicular to $\nabla\chi$. If all three components of this equation were satisfied simultaneously (which, in general, is not possible as noted before), then the original full MHD momentum equation would also be satisfied and momentum would be conserved exactly. We can estimate the magnitude of momentum conservation error by considering the components of the vorticity-like equation perpendicular to \(\nabla\chi\). This equation can be written as
\begin{equation}
	\label{eq:oldvle}
	\begin{aligned}
		&\tderiv{\st{\vec\omega}} + \mu_0 v_\parallel\tderiv{\vec j} + \nabla v_\parallel\times\tderiv{\vec B} - \frac{\mu_0\vec j}{B_v}\llderiv v^2 - \nabla\left(\frac{\llderiv v^2}{B_v}\right)\times\vec B + \nabla\times(\vec\omega\times\vec v) \\
		&- \frac{\mu_0\vec j}{B_v^2}\nabla\chi\cdot(\vec\omega\times\vec v) - \nabla\left[\frac{\nabla\chi\cdot(\vec\omega\times\vec v)}{B_v^2}\right]\times\vec B + \st{\vec\omega}~\frac{P}{\rho} + \nabla\left(\frac{P}{\rho}\right)\times\st{\vec v} = \frac{\vec B}{\rho^2}(\vec j\cdot\nabla\rho) \\
		&- \frac{\vec j}{\rho^2}(\vec B\cdot\nabla\rho) + \frac{1}{\rho}(\vec B\cdot\nabla)\vec j - \frac{1}{\rho}(\vec j\cdot\nabla)\vec B - \frac{\vec j}{\rho B_v^2}\nabla\chi\cdot(\vec j\times\vec B) - \nabla\left[\frac{\nabla\chi\cdot(\vec j\times\vec B)}{\rho B_v^2}\right]\times\vec B \\
		&+ \frac{1}{\rho^2}\nabla\rho\times\nabla p + \frac{\vec j}{\rho B_v}\llderiv p + \nabla\left(\frac{\llderiv p}{\rho B_v}\right)\times\vec B,
	\end{aligned}
\end{equation}
where the reduced velocity $\st{\vec v} = -\nabla\chi\times(\vec e_\chi\times\vec v) = \nabla\Phi\times\nabla\chi/B_v^2$ and the reduced vorticity $\st{\vec\omega} = \nabla\times\st{\vec v}$ were introduced, and the viscosity term is not considered. If the components of this equation perpendicular to $\nabla\chi$ are identically zero, then there is no approximation in the reduction as nothing is being neglected, and momentum is still conserved. The most general case in which these components are zero is the following:
\begin{equation}
\label{eq:exact}
\llderiv u = 0, \quad u\in\{g^{ik}, \Phi, \Psi, \vpar, p, \rho, P\}
\end{equation}
where $g^{ik}$ are the components of the metric tensor of the vacuum field-aligned coordinate system. As can be shown by a simple calculation, in this case both $\st{\vec\omega}$ and $\vec j$ will be directed strictly along $\nabla\chi$. If we allow either $g^{ik}$, $\Phi$ or $\Psi$ to vary along $\nabla\chi$, the same calculation will show that $\st{\vec\omega}$ has nonzero perpendicular components. This will cause $\partial\st{\vec\omega}/\partial t$ to have nonzero components perpendicular to $\nabla\chi$, which cannot be canceled by any other terms since there are no more time derivatives involving $\Phi$ in the equation. Similarly, if we let any of the other quantities vary along $\nabla\chi$, the last term and the seventh term on the RHS (pressure), the first term and the seventh term on the RHS (density), the last term on the LHS ($P$) and third term on the LHS ($\vpar$) will have nonzero perpendicular components, which will not be canceled by any other terms. If the conditions \eqref{eq:exact} are met, then only the sixth and eighth terms on the LHS have nonzero perpendicular components. As can be shown by a simple expansion of the eighth term:
\begin{equation*}
[\nabla\times(\vec\omega\times\vec v)]^\perp - \left[\nabla\left[\frac{\nabla\chi\cdot(\vec\omega\times\vec v)}{B_v^2}\right]\times\vec B\right]^\perp \equiv 0.
\end{equation*}

Now we consider the new reduced model. For the first projection operator \eqref{eq:newprojop}, the vorticity-like equation analogous to equation \eqref{eq:oldvle} can be written as:
\begin{equation}
	\label{eq:newvle}
	\begin{aligned}
		&\nabla\left(\frac{\rho}{B_v^2}\right)\times\tderiv{\st{\vec v}} + \frac{\rho}{B_v^2}\tderiv{\st{\vec\omega}} + \nabla\left(\frac{\rho}{B_v^2}\tderiv{\vpar}\right)\times\vec B + \frac{\mu_0\rho}{B_v^2}\tderiv{\vpar}\vec j + \nabla\left(\frac{\rho\vpar}{B_v^2}\right)\times\tderiv{\vec B} + \frac{\mu_0\rho\vpar}{B_v^2}\tderiv{\vec j} \\
		&+ \frac{1}{2}\nabla\left(\frac{\rho}{B_v^2}\right)\times\nabla v^2 + \nabla\times\left(\frac{\rho\vec\omega\times\vec v}{B_v^2}\right) + \nabla\left(\frac{P}{B_v^2}\right)\times\st{\vec v} + \frac{P}{B_v^2}\st{\vec\omega} + \nabla\left(\frac{P\vpar}{B_v^2}\right)\times\vec B \\
		&+ \frac{\mu_0 P\vpar}{B_v^2}\vec j = \frac{2\vec B}{B_v^3}\vec j\cdot\nabla B_v - \frac{2\vec j}{B_v^3}\vec B\cdot\nabla B_v + \frac{1}{B_v^2}(\vec B\cdot\nabla)\vec j - \frac{1}{B_v^2}(\vec j\cdot\nabla)\vec B + \frac{2}{B_v^3}\nabla B_v\times\nabla p.
	\end{aligned}
\end{equation}
Equation \eqref{eq:phieq} is just the contravariant $\chi$ component of this equation. In the above equation, the third and eleventh terms on the LHS will have perpendicular components even when condition \eqref{eq:exact} is met; namely, $\pgrad(\rho B_v^{-2}\partial\vpar/\partial t)\times\nabla\chi$ from the third term and $\pgrad(P\vpar B_v^{-2})\times\nabla\chi$ from the eleventh term. In addition, while the eighth term on the LHS of equation \eqref{eq:newvle} is analogous to the sixth term on the LHS of equation \eqref{eq:oldvle}, there is no term in equation \eqref{eq:newvle} analogous to the eighth term of equation \eqref{eq:oldvle} to cancel the nonzero perpendicular components of its eighth term. Thus, a more stringent condition is required to ensure that the perpendicular components of equation \eqref{eq:newvle} are identically zero and momentum is exactly conserved in the new reduced MHD model. Namely, in addition to the conditions \eqref{eq:exact}, we must also have $\vpar = 0$. This additional condition causes the third and eleventh terms to vanish, and the eighth term becomes:
\begin{equation*}
	\nabla\times\left(\frac{\rho\st{\vec\omega}\times\st{\vec v}}{B_v^2}\right) = \nabla\times\left(\frac{\rho\st{\omega}_\chi\pgrad\Phi}{B_v^2}\right) = \nabla\left(\frac{\rho\st{\omega}_\chi}{B_v^2}\right)\times\pgrad\Phi - \frac{\rho\st{\omega}_\chi}{B_v^2}\nabla\left(\frac{\llderiv\Phi}{B_v}\right)\times\nabla\chi.
\end{equation*}
Clearly, if condition \eqref{eq:exact} is met, the eighth term, as shown above, will not have any perpendicular components.

As can be seen, the original reduced model \citep{nikulsin2019a} has better momentum conservation properties than the new reduced model (section \ref{sec:changes}). This was the reason why we initially attempted to work with the original model. However, as energy conservation is generally more important than momentum conservation, we will use the new model for future simulations.

\subsection{Global momentum conservation error (numerical examples)}

\begin{figure}
	\includegraphics[scale=0.9]{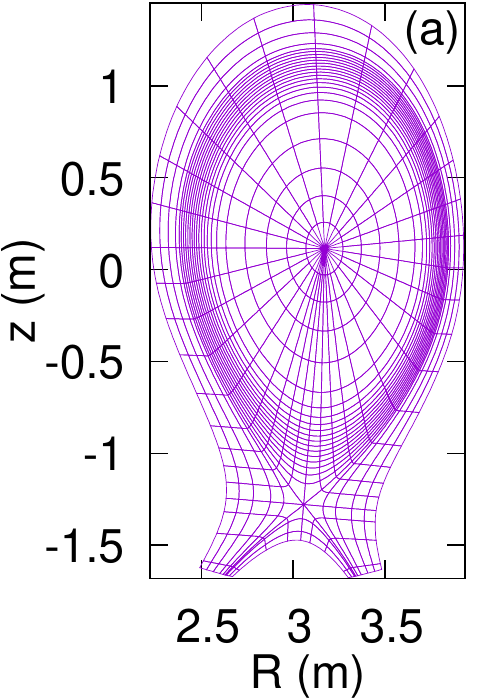}\hfill
	\includegraphics[scale=0.3]{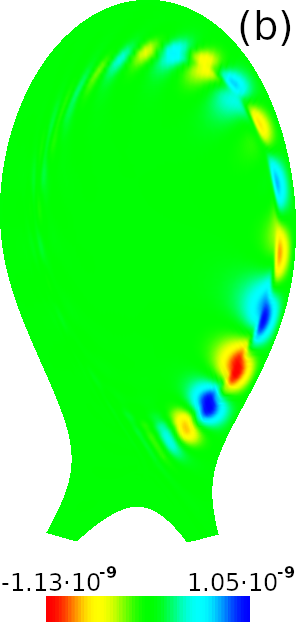}
	\caption{A flux-aligned grid used for simulating the ballooning mode, shown here with reduced resolution for clarity (a), and the sum of $n>0$ Fourier mode of $F_0\Psi$ (JOREK units) in the standard tokamak model at 250 time steps, at $t=641.1$ Alfv\'en times (b).}
	\label{fig:ballooning}
\end{figure}

As a test case, we will use a simple ballooning mode in an X-point plasma in a tokamak with an aspect ratio of 3 (Figure \ref{fig:ballooning}). In the simulations shown here only the Fourier modes with $0 \leq n \leq 6$ were included. The simulation parameters were as follows: $F_0 = 3~\mathrm{T\cdot m}$, a time step of 3 Alfv\'en times, a resistivity of $\eta = 3.8764\cdot 10^{-6}~\mathrm{\Omega\cdot m}$ ($2\cdot 10^{-6}$ JOREK units), and a viscosity of $\mu = 2.293\cdot 10^{-7}~\mathrm{kg/(m\cdot s)}$, corresponding to $\mu_\perp^t = 4\cdot 10^{-6}$ JOREK units. The initial conditions were set up by solving the Grad-Shafranov equation with
\begin{equation*}
    FF'(\psi_n) = \frac{1}{2}\left[1.6(1 - \psi_n) - 0.43\cosh^{-2}\frac{\psi_n - 0.9}{0.07}\right]\left(1 - \tanh\frac{\psi_n - 1}{0.03}\right),
\end{equation*}
$T(\psi_n) = 0.015(1 - 0.66\psi_n)[1 - \tanh((\psi_n - 0.94)/0.08)]/2 + 3\cdot 10^{-4}$, and \\${\rho(\psi_n) = [1 - \tanh((\psi_n - 0.94)/0.08)]/2 + 0.01}$, where $FF'$ is in units of $\mathrm{T^2\cdot m}$, $T$ is in $(10^{20}k_B\mu_0)\cdot\mathrm{K}$ and $\rho$ is in $3.346\cdot 10^{-7}\mathrm{kg/m^3}$. In addition, a hyperresistivity of $\eta_h = 5.8146\cdot 10^{-10}~\mathrm{\Omega\cdot m^2}$ ($3\cdot 10^{-10}$ JOREK units) was used in the simulations. We compare the new reduced model with $\vpar = 0$ to the new reduced model with $\vpar\neq 0$. Since in the tokamak limit these models match the standard JOREK tokamak models, we used the tokamak models for the actual simulations.

To quantify the momentum conservation error, we calculate the total linear momentum in the Cartesian $x$- and $y$-directions for the entire plasma. Physically, this momentum should be zero throughout the simulation, however, due to the inaccuracies discussed in the previous subsection, a nonzero momentum tends to appear. Here, we choose the $x$-direction as the $R$-direction when $\phi=0$ and the $y$-direction as the $R$-direction when $\phi=\pi/2$, with the $z$-axis being the axis of symmetry of the torus. We do not consider momentum in the $z$-direction since the global $z$-momentum is conserved in the tokamak limit, as can be seen by setting $\Phi^* = \ln R$ in equation \eqref{eq:pophi} and recalling that $B_v = F_0/R$ in the tokamak case.

\begin{figure}
	\includegraphics[scale=0.58]{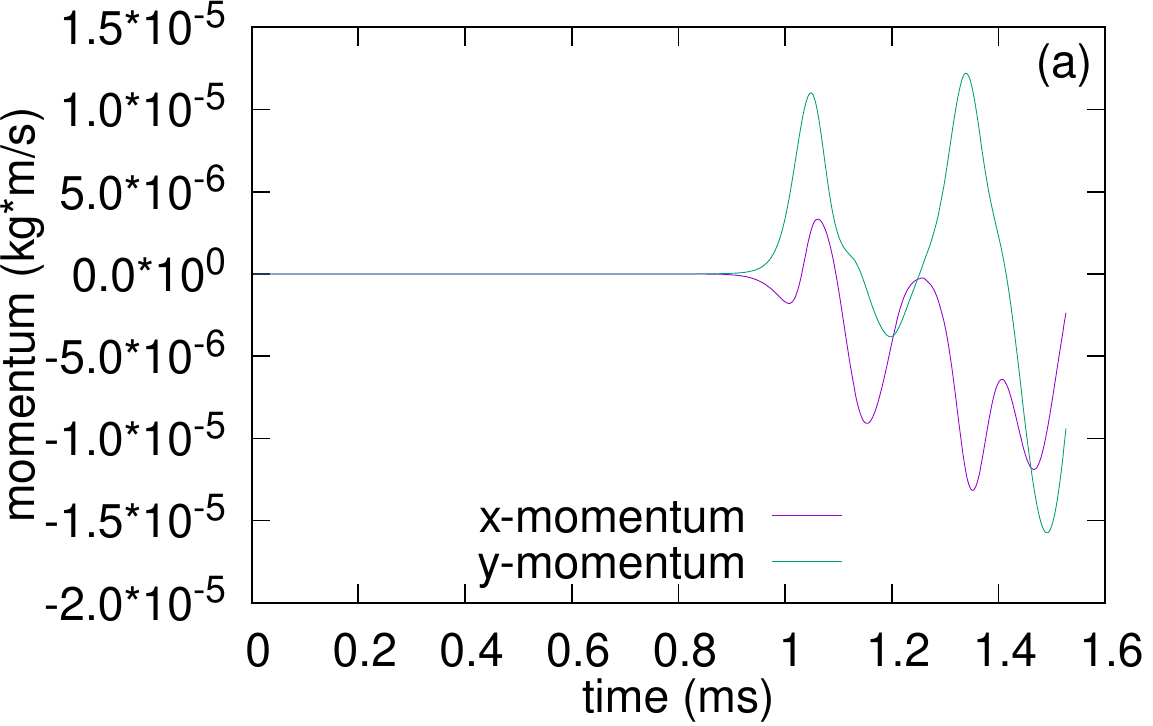}
	\includegraphics[scale=0.58]{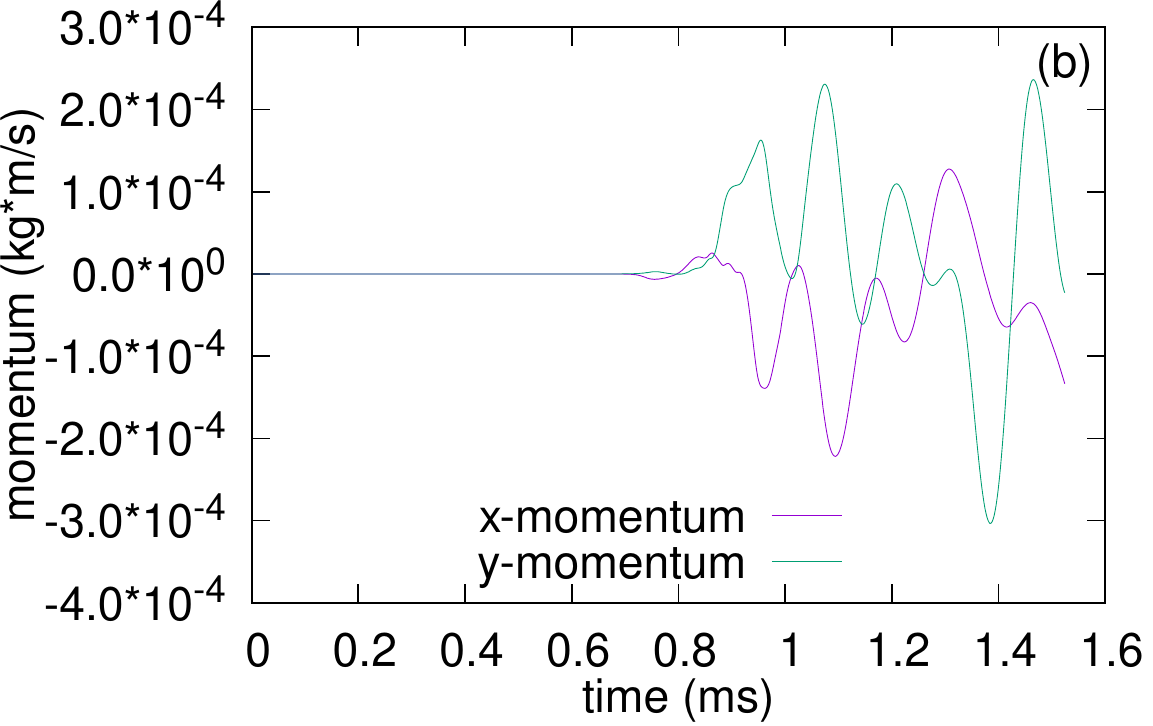}
	\caption{Total linear momentum in the Cartesian $x$- and $y$-directions as a function of time. Momentum evolution is shown for simulations using the reduced model with $\vpar=0$ (a), and the reduced model with $\vpar\neq 0$ (b).}
	\label{fig:momentum}
\end{figure}

As can be seen in Figure \ref{fig:momentum}, after about 1.5~ms, the momentum conservation error in the model with $\vpar\neq 0$ is worse by more than an order of magnitude than that in the model with $\vpar = 0$. In both cases, the instability has reached nonlinear saturation, and at that point the momenta no longer grow in amplitude, but tend to oscillate around zero with the amplitude of the oscillations remaining at the same order of magnitude.

\section{Conclusion}

In continuation of our previous work \citep{nikulsin2019a}, we have implemented the stellarator-capable reduced MHD model derived in our previous paper and tested it in the tokamak limit. We find that, even in the simple case of a tearing mode in a circular high aspect ratio tokamak, energy conservation can only be approximately satisfied, due to the possibility of nonphysical gain or loss of kinetic energy. Our original plan, as presented in \citet{nikulsin2019a}, was to use the full energy conservation equation when evolving pressure, thus ensuring energy conservation. However, that would require the internal energy to be much larger than the kinetic energy, thus excluding low $\beta$ situations. Even without exact energy conservation, meaningful linear results, such as growth rates, could be obtained without any further modifications, however the lack of energy conservation required artificial dissipation to be introduced to prevent a crash after nonlinear saturation was reached due to nonphysical kinetic energy buildup. Introducing artificial dissipation in the form of a hyperviscosity term kept the kinetic energies in check and allowed the simulations to proceed into the post-saturation regime. Although exact energy conservation could not be achieved, the error in energy conservation remained small because the kinetic energy is small compared to the total energy, and the magnetic energy evolved correctly. Thus, energy should be approximately conserved whenever the kinetic energy is small, which is usually the case in fusion-relevant situations. We also found that the kinetic energy blowup was triggered by numerical errors from the magnetic field. Replacing the magnetic field equation with a different form that is analytically equivalent to the original, but more numerically stable alleviates the need for hyperviscosity, although the energy conservation error in this case is larger than when hyperviscosity is present.

To remedy the lack of energy conservation in our original model, we used a new set of projection operators to obtain scalar momentum equations from the full MHD vector momentum equation. With these new projection operators, the reduced MHD model matches the standard JOREK reduced MHD models for tokamaks \citep{franck2015energy,hoelzl2020the} in the tokamak limit, i.e. when $\chi = F_0\phi$, where $\nabla\chi$ is the vacuum magnetic field, and $\phi$ is the toroidal angle. When using the new model, energy is conserved almost exactly, with a small amount of error due to temporal discretization and the use of a truncated Fourier decomposition in the toroidal direction. However, the new model also has more error in the momentum conservation than the original model whenever $\vpar\neq 0$.\\

The authors thank Rohan Ramasamy, Florian Hindenlang, Boniface Nkonga, Guido Huijsmans, Vinodh Bandaru and Erika Strumberger for fruitful discussions.

This work has been carried out within the framework of the EUROfusion Consortium and has received funding from the Euratom research and training program 2014-2018 and 2019-2020 under grant agreement No. 633053. The views and opinions expressed herein do not necessarily reflect those of the European Commission.

\appendix
\section{}\label{sec:appendix}

In most reduced MHD models used for tokamaks, including the tokamak reduced MHD model implemented in JOREK \citep{franck2015energy,hoelzl2020the}, field compression is neglected altogether via the ansatz $\vec B = F_0\nabla\phi + \nabla\Psi\times\nabla\phi$, which assumes that the background vacuum field is purely toroidal and only allows field line bending corrections to that. This is possible due to a property of the projection operator used, ${\nabla\phi\cdot\nabla\times(R^2*}$, which allows exact force balance in the reduced system of equations when equilibria satisfy the Grad-Shafranov equation, even if field compression is neglected. Consider the second (poloidal momentum) equation in section 2.6 of \citet{franck2015energy} with $u = v_\parallel = 0$\footnote{We include here a factor of $1/\mu_0$ in the second and third terms, which was not present in \citet{franck2015energy} due to the current being normalized.}:
\begin{equation}
	0 = -\frac{1}{R}[R^2,p] + \frac{1}{R\mu_0}[\Psi,\st j] - \frac{F_0}{R^2\mu_0}\frac{\partial\st j}{\partial\phi},\label{eq:redeq}
\end{equation}
where $R$ is the distance from the axis of symmetry (major radius coordinate), $p$ is pressure, $\Psi$ is the poloidal flux and $\st j = \Delta^*\Psi$. One can easily see that this equation can be obtained by applying the projection operator $\nabla\phi\cdot\nabla\times(R^2$ to the static equilibrium condition $\vec j\times\vec B = \nabla p$. Under the assumption of axisymmetry, the derivative in the last term in equation \eqref{eq:redeq} is zero. In addition, since pressure is a flux function, we have $\nabla p = p'\nabla\Psi$. Thus, equation \eqref{eq:redeq} becomes
\begin{equation}
	0 = \mu_0 p'[R^2,\Psi] + [\Delta^*\Psi,\Psi].\label{eq:redeq2}
\end{equation}
Meanwhile, the Grad-Shafranov equation reads:
\begin{equation}
	\Delta^*\Psi = -\mu_0 R^2 p' - FF'.
\end{equation}
Substituting this into the second term in equation \eqref{eq:redeq2}, we have
\begin{equation*}
	0 = \mu_0 p'[R^2,\Psi] - \mu_0[R^2 p',\Psi] - [FF',\Psi] \equiv - \mu_0 R^2 [p',\Psi] - [FF',\Psi] \equiv 0,
\end{equation*}
where the last two terms are identically zero since $p'$ and $FF'$ are both functions of only $\Psi$, so their Poisson bracket with $\Psi$ must be zero. Finally, the last (parallel momentum) equation in section 2.6 of \citet{franck2015energy} with $u = v_\parallel = 0$ reads:
\begin{equation}
	0 = -\frac{1}{R}[p,\Psi] - \frac{F_0}{R^2}\frac{\partial p}{\partial\phi},
\end{equation}
and can be obtained by projecting $\vec j\times\vec B = \nabla p$ onto $\vec B$. In this equation, the first term is again identically zero due to $p$ being a flux function, and the second term is zero due to axisymmetry. Thus, any solution of the Grad-Shafranov equation is also an exact solution of the reduced MHD system with $\vec v = 0$, and so this version of reduced MHD has the same set of axisymmetric equilibria as full MHD. 

As is clear in the proof above, the Grad-Shafranov equation plays a crucial role. Without it, the reduced MHD system would not admit a solution with $\vec v = 0$ even if we were to find a $(\Psi,p)$ satisfying $\vec j\times\vec B = \nabla p$. In other words, for three dimensional equilibria, one should not expect to have force balance in the reduced MHD system. While not proven, it seems unlikely to us that a different choice of projection operator could resolve this problem.

At high aspect ratios, lack of force balance becomes less significant due to the neglected field compression becoming smaller, and vanishing completely in the cylindrical limit. However, for an aspect ratio of 10, the lack of force balance is already significant. In order to carry out the simulations presented in section \ref{sec:examples} with our original model, we had to introduce a static force balancing term, which was not considered in \citet{nikulsin2019a}. For the sake of clarity, we consider the force balancing term before the projection operator is applied, however note that the full MHD vector momentum equation is overconstrained if the velocity ansatz is used but the projection operator is not applied. The modified equation is as follows:
\begin{equation}
	\label{eq:nsfbt}
	\tderiv{}(\rho\vec v) + \nabla\cdot(\rho\vec v\vec v) = \vec j\times\vec B - \nabla p + \rho\nu\Delta\vec v + (\vec j_f\times\vec B_f - \vec j\times\vec B)|_{t=0},
\end{equation}
where $\vec B$ is the magnetic field given by the ansatz, $\mu_0\vec j = \nabla\times\vec B$ and $\vec B_f$ and $\vec j_f$ are the full MHD magnetic field and current, i.e. the actual field and current, not their ansatz-based approximations. At $t=0$, the two ansatz-based Lorentz force terms cancel, and the forces are balanced by the full MHD Lorentz force; the first Lorentz force term then evolves with time while the force balancing term remains frozen. Such an approach should work in most cases, as instabilities tend to not compress the magnetic field due to the large amount of energy required to do so. In the tokamak limit, the force balancing term reduces to $-FF'~\nabla\Psi|_{t=0}/R^2$.

We can show that this extra term does not violate global conservation of momentum. If we ignore the viscosity term, equation \eqref{eq:nsfbt} can be rewritten as
\begin{equation}
	\tderiv{}(\rho\vec v) + \nabla\cdot\left[\rho\vec v\vec v + \left(p + \frac{B^2 + (B_f^2 - B^2)|_{t=0}}{2\mu_0}\right)\overleftrightarrow{I} - \frac{\vec B\vec B + (\vec B_f\vec B_f - \vec B\vec B)|_{t=0}}{\mu_0}\right] = 0.
\end{equation}
Integrating over the plasma volume and assuming that both the reduced and full magnetic fields are tangential to the plasma boundary, we obtain
\begin{equation}
	\frac{d}{dt}\int_V \rho\vec v dV = -\oint_{\partial V} \left(p + \frac{B^2 + (B_f^2 - B^2)|_{t=0}}{2\mu_0}\right)d\vec S.
\end{equation}
The RHS is just the force exerted by the walls on the plasma. Note that the term $(B_f^2 - B^2)|_{t=0}/(2\mu_0)$ is approximately the energy density stored by compressing the background vacuum field in equilibrium. Since instabilities tend to not compress the magnetic field, the energy density stored in the compression should be roughly constant throughout the simulation, we should have $B^2 + (B_f^2 - B^2)|_{t=0} \approx B_f^2$, where $B_f^2/(2\mu_0)$ is the total magnetic energy density, which includes both field line bending and compression.

To conclude, we note that the momentum conservation properties discussed in section \ref{sec:momentum} remain unchanged with the addition of the force balancing term. To show this, we substitute $p = p|_{t=0} + \widetilde{p}$ into equation \eqref{eq:nsfbt}, where $\widetilde{p}$ is defined as ${p - p|_{t=0}}$. Then, because $(\vec j_f\times\vec B_f)|_{t=0} = \nabla p|_{t=0}$, those terms will cancel, and the only term remaining in equation \eqref{eq:nsfbt} that was not originally present in the vector momentum equation \eqref{eq:vrmhd} is $-(\vec j\times\vec B)|_{t=0}$. However, this is just the reduced Lorentz force term at $t=0$, which has the same structure as the time varying reduced Lorentz force term that we have already dealt with. Most importantly, if $\llderiv\Psi = \llderiv g^{ik} = 0$, then $\vec j\parallel\nabla\chi$, where $g^{ik}$ are the components of the metric tensor of the vacuum field-aligned coordinate system. The conditions for exact momentum conservation are thus
\begin{equation}
	\llderiv u = 0, \qquad u\in\{g^{ik}, \Phi, \Psi, v_\parallel, \widetilde{p}, \rho, P\}.
\end{equation}
The only minor difference from condition \eqref{eq:exact} is the replacement of $p$ with $\widetilde{p}$.

\bibliographystyle{jpp}

\bibliography{references}

\end{document}